# The transitional kinetics between open and closed Rep structures can be tuned by salt via two intermediate states

Jamieson A L Howard[1], Benjamin Ambrose[2,3], Mahmoud A S Abdelhamid[2], Lewis Frame[1], Antoinette Alevropoulos-Borrill[1], Ayesha Ejaz[4], Lara Dresser[1], Maria Dienerowitz[5], Steven D Quinn[1,6], Allison H Squires[7,8], Agnes Noy[1,6], Timothy D Craggs[2], and Mark C Leake[1,6,9] [†]

[1] School of Physics, Engineering and Technology, University of York, York, YO10 5DD, UK

[2] Department of Chemistry, University of Sheffield, Sheffield S3 7HF, U.K.

[3] Current address: Single Molecule Imaging Group, MRC-Laboratory of Medical Sciences, London, W12 0HS, UK.

[4] Department of Chemistry, University of Chicago, Chicago, IL, USA.

[5] SciTec Department, Ernst-Abbe-Hochschule, University of Applied Sciences, Jena, Germany.

[6] York Biomedical Research Institute, University of York, York, YO10 5DD, UK.

[7] Pritzker School of Molecular Engineering, University of Chicago, Chicago, IL, USA.

[8] Institute for Biophysical Dynamics, University of Chicago, Chicago, IL, USA.

[9] Department of Biology, University of York, York, YO10 5DD, UK.

[†] For correspondence. Email mark.leake@york.ac.uk

**Abstract**

DNA helicases undergo conformational changes; however, their structural dynamics are poorly understood. Here, we study single molecules of superfamily 1A DNA helicase Rep, which undergo conformational transitions during bacterial DNA replication, repair and recombination. We use time-correlated single-photon counting (TCSPC), fluorescence correlation spectroscopy (FCS), rapid single-molecule Förster resonance energy transfer (smFRET), Anti-Brownian ELectrokinetic (ABEL) trapping and molecular dynamics simulations (MDS) to provide unparalleled temporal and spatial resolution of Rep's domain movements. We detect four states revealing two hitherto hidden intermediates (S2, S3), between the open (S1) and closed (S4) structures, whose stability is salt dependent. Rep's open-to-closed switch involves multiple changes to all four subdomains 1A, 1B, 2A and 2B along the S1→S2→S3→S4 transitional pathway comprising an initial truncated swing of 2B which then rolls across the 1B surface, following by combined rotations of 1B, 2A and 2B. High forward and reverse rates for S1→S2 suggest that 1B may act to frustrate 2B movement to prevent premature Rep closure in the absence of DNA. These observations support a more general binding model for accessory DNA helicases that utilises conformational plasticity to explore a multiplicity of structures whose landscape can be tuned by salt prior to locking-in upon DNA binding.

**Introduction**

DNA helicases are ubiquitous molecular motors of nucleic acid metabolism that play crucial roles in DNA repair and the regulation and maintenance of genetic stability. They utilize chemical potential energy derived from nucleoside triphosphate (NTP) hydrolysis usually in the form of ATP to translocate along a nucleic acid strand in a unidirectional manner[1–4] normally by 1-2 base pairs per NTP hydrolysed[1,2], and disrupt hydrogen bonds between complementary base pairs. Defects in viral/bacterial and human DNA helicases are linked to infectious diseases, genetic disorders and cancers[5] while their crucial roles have made them an attractive target for several therapeutic interventions[5–7]. They are divided into six superfamilies SF1-SF6 based on shared sequence motifs[1] with processive DNA helicases from SF1[8] and SF2[9] involved in virtually all aspects of RNA and DNA metabolism. SF1A is a sub-division of SF1 that includes several well-characterized DNA helicases that aid DNA replication machinery in progressing through nucleoprotein complexes which would otherwise act as a replication block[10–15]. *Escherichia coli* protein Rep is an ATP-dependent accessory replicative DNA helicase of the superfamily 1A subdivision of SF1 that is essential under rapid growth conditions[11,12,16], and is structurally comparable to *E. coli* UvrD and *Geobacillus stearothermophilus* PcrA. Rep contains four subdomains (1A, 2A, 1B and 2B), in which 2B appears capable of undergoing a substantial rotation about a hinge connected to 2A[1,17–23]. As with all superfamily 1A DNA helicases, Rep has 3'-5' directionality and as such likely operates on the leading strand template displacing proteins ahead of the replication fork[11,12]. This activity is aided through a physical interaction via the C-terminus with the 5'-3' replicative helicase DnaB, which targets Rep to the replication fork[11,24–27] whilst also acting ahead of the bacterial replisome.

Monomeric Rep is a processive translocase on single-stranded DNA (ssDNA), but not a functional DNA helicase in that it appears unable to unwind duplex DNA in the absence of any other monomeric Rep molecules acting on the same duplex[17,28–31]. Monomeric unwinding of double-stranded DNA (dsDNA) is activated in the RepΔ2B mutant in which the 2B subdomain is removed[28,32,33]; this points towards an autoinhibitory function possibly regulating spatiotemporal activity[28,34]. However, removal of 2B also prevents displacement of the protein component of nucleoprotein complexes by the DNA helicase[33] and *repΔ2B* is unable to complement for *Δrep in vivo*[32,33]. The importance of the 2B conformation on Rep activity agrees with its homologues UvrD in *E. coli* and PcrA in *Geobacillus stearothermophilus* in which 2B can rotate between 130° and 160° relative to the other three subdomains. In these homologues, the closed conformation confers a greater processivity and level of unwinding than in the open form[2,4,17–20,22–24,35,36]. Similarly, Rep that has been crosslinked in the closed state acts as a "super helicase" in terms of processivity and force generated[20]. Although wild type monomers of superfamily 1A DNA helicases appear unable to unwind significant lengths of dsDNA, this activity is activated when multiple DNA helicases act on the same DNA molecule[23,29,37–40] or when activated by another co-factor as in the case of UvrD and MutL and Ku (in *M. tuberculosis*) or PcrA and RepD[41–44] and generally correlates with the 2B subdomain being pushed towards a more "closed" state. Recently, the application of single-molecule Förster resonance energy transfer (smFRET) experiments with 32 ms time-resolution, suggested that the 2B subdomain in immobilised UvrD helicases populates as least four interconverting rotational conformational states whether bound or not to its DNA substrate[28]. However, whether this feature is a characteristic of other 1A superfamily DNA helicases, and if this structural heterogeneity exists in the absence of the DNA substrate, remains unclear.

Here, we have used time-correlated single photon counting (TCSPC), fluorescence correlation spectroscopy (FCS), time-resolved smFRET, Anti-Brownian ELectrokinetic (ABEL) trapping and total internal reflection fluorescence (TIRF) microscopy and integrated these high-precision data with all-atom molecular dynamics simulations (MDS), to study freely diffusing and immobilised Rep molecules. We find that in both the absence and presence of its ssDNA substrate, Rep exhibits four states, S1 (low FRET) and S4 (high FRET) corresponding to the open and closed structures previously identified from X-ray crystallography[18], plus two previously undiscovered intermediate FRET states S2 and S3. We observe that changing the NaCl concentration from high (500 mM) through the intermediate physiological (150 mM) to low (down to 10 mM) biases the states from open to closed conformations. Using photon-by-photon hidden Markov modelling (H2MM) coupled to MDS and structure interpolation we find that the most likely transitional pathway involves S1→S2→S3→S4 comprising several discrete movements of subdomains 1B, 2A and 2B that involves an initial truncated 2B swing and then roll across the 1B surface followed by combined rotations of 1B, 2A and 2B. These structural predictions also indicate that, for the S1→S2 transition, 1B may act to sterically hinder 2B movement with a result that the likelihood for premature closing of Rep in the absence of DNA is reduced. A similar recent study on the accessory DNA helicase UvrD, a homologue to Rep, also indicated the presence of four states between UvrD's open and closed conformations, but suggested that its closure involves a single 2D swing-arm mechanism of UvrD's equivalent 2B subdomain[19].

We find that five out of the 12 transitions which are in principle possible between any pair of these four Rep states show a strong rate dependence on NaCl concentration. Simulations indicate that, upon DNA binding, the extent of Rep conformational dynamics is significantly reduced. Our findings point towards an exploratory conformational strategy for Rep in which a range of structural states can be reversibly trialled, steered by the salt concentration dependence, prior to locking-in on a more rigid conformation upon binding to DNA. Our observations, which have been primarily enabled through a diverse range highly integrated and innovative biophysical experiments and computational simulations, pave the way for usefully exploring general conformational dynamics of other DNA helicases in response to accessory factors, and for further elucidating the mechanism by which the protein component of nucleoprotein complexes are displaced to ensure genomic integrity. Such combined investigations may help to reveal wider strategies for salt-mediated regulation of enzyme-substrate binding where multiple conformational states of the enzyme are involved.

**Results**

**Changing salt concentration indicates signs of conformational variability for Rep using ensemble TCSPC and FCS**

To investigate the conformational states of free Rep we generated a Rep Δcysteine mutant from the wild type gene and used this to create a Rep two cysteine mutant by mutating alanine residues 97 and 473 located on subdomains 1B and 2B respectively to non-native cysteine (Methods). This mutant was cloned into *E. coli* and characterised for functionality using complementation assays (Supplementary Fig. 1, Supplementary Table 1 and Methods)[11,45]. Following overexpression and purification, we incorporated fluorescent dyes onto the cysteine sites using a preliminary range of candidate FRET donor/acceptor pairs and concentrations, with donor Alexa Fluor 546 and acceptor Alexa Fluor 647 yielding the best response and smallest propensity to aggregation for the maximum number of FRET-active species (Supplementary Fig. 2) to report on conformational closing as suggested from static X-ray crystallographic structures of open and closed states[18] (Fig. 1a). Using absorption measurements, we estimate that there were approximately two dyes per Rep molecule as judged by the concentration of dye molecules compared to the concentration of protein, with each dye present in equimolar amounts. Using the smFRET in-solution data we estimate that this corresponds to 8% for doubly labelled donor-acceptor from the total of all fluorescently labelled Rep; this is in line with the labelling efficiency levels reported recently by others on Rep[46]. We confirmed that this labelled mutant retained DNA helicase function by measuring its ability to unwind DNA *in vitro* relative to the unlabelled wild type protein. (Fig. 1b). Phosphoradioimaging of radiolabelled DNA (Methods) indicates that labelled A97CA473C Rep unwound 74.6 ± 4.7% (±SD) of forked DNA compared to wild type 71.7 ± 1.3%, consistent with retention of activity. Fluorescence-based timecourse assays (Methods) in which an Alexa Fluor 488 dye is unquenched upon unwinding of a DNA substrate, also indicated that fluorescently labelled Rep retains wild type activity to within experimental error over the assay's full duration of 1 hour (Fig. 1c), with comparable initial velocities (5.97 ± 0.34 vs 5.44 ± 0.47 a.u. s$^{-1}$) and $k_{obvs}$ (299 ± 17 vs 272 ± 23 a.u. µM$^{-1}$ s$^{-1}$) for wild type and labelled Rep respectively.

We first investigated the doubly labelled Rep in solution using ensemble FRET spectroscopy through excitation of the Alexa Fluor 546 donor dye via TCSPC to measure its fluorescence lifetime. FRET between the donor and acceptor is inversely proportional to the 6$^{th}$ power of inter-dye distance, such that closer proximity leads to donor quenching manifested as a distance-dependent shortening of the donor fluorescence lifetime[47]. We measured the lifetime in salt buffers corresponding to 10 mM (low), 150 mM (intermediate) and 500 mM (high) NaCl, determined these from fitting of the fluorescence intensity decay profiles with a sum of exponential components (Methods). At low salt, the amplitude-weighted average fluorescence lifetime ($\tau_{av}$) was 3.51 ns, whereas for intermediate and high salt conditions $\tau_{av}$ increased to 3.57 ns and 3.61 ns, respectively, indicative of a progressive increase in the inter-dye distance corresponding to a more open conformation. Based on free Alexa Fluor 546 dye controls under identical buffer conditions (Supplementary Fig. 3), this increase in lifetime was not attributable to photophysical artifacts associated with the free dye. However, when similar experiments were performed with Rep labelled only with Alexa Fluor 546, we measured an initial ensemble lifetime of approximately 1.8 ns that increased to 3.6 ns upon addition of NaCl. This suggests that the donor also exhibits modest and reproducible NaCl-dependent fluorescence shifts, which could be attributed to conformational flexibility, possibly via a mechanism involving protein-induced fluorescence enhancements (PIFE)[48–50]. The smallest number of fitting parameters required to describe the fluorescence decay of Alexa Fluor 546 on Rep was three (Supplementary Table 2), hinting at the presence of multiple Rep conformations; however, ground-state dye variability, the propensity for sample aggregation, and intrinsic fluctuations in molecular structure all potentially contributed to fluorescence decay complexity.

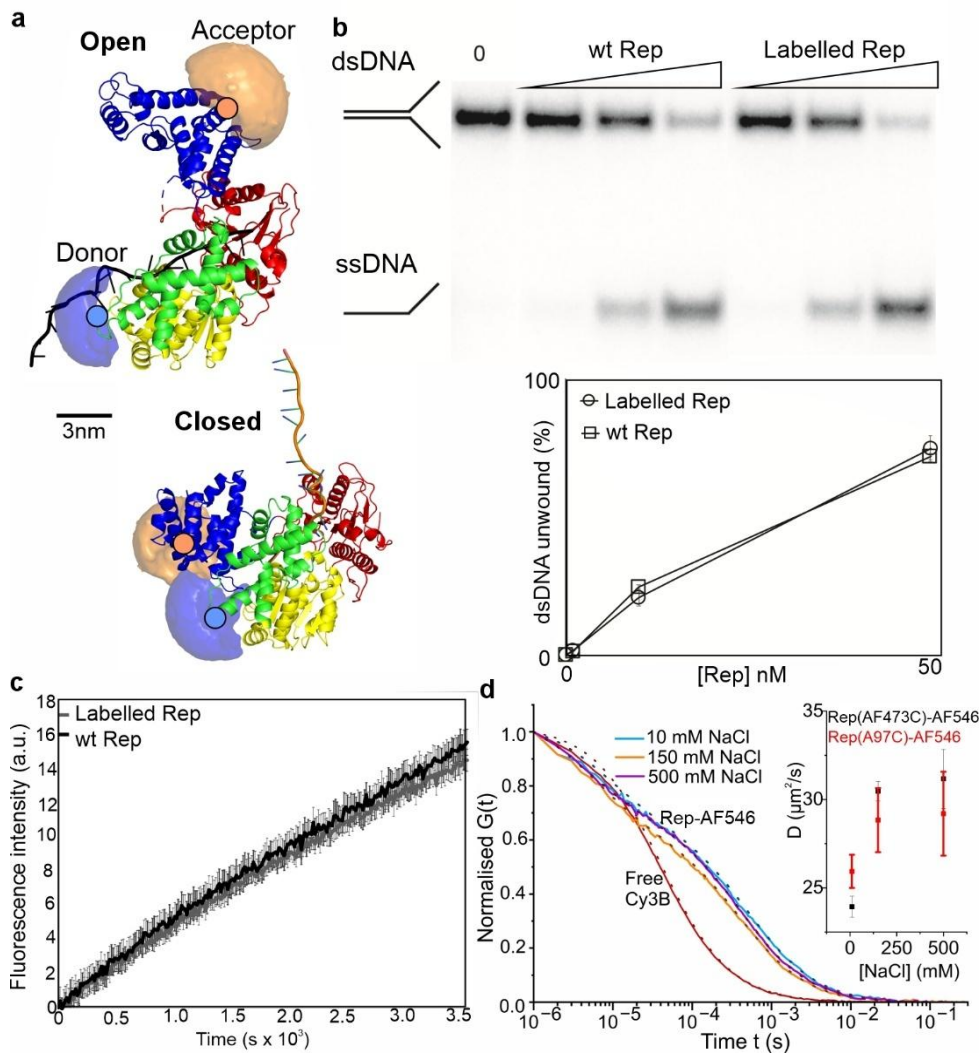

**Figure 1. FRET pairs on Rep report on its conformational state in solution. a.** Crystal structures of Rep occupy both open and closed conformations; the separate subdomains are coloured yellow (1A), green (1B), red (2A), blue (2B). Donor and acceptor dyes here labelled at A97C and A473C (orange and blue filled circles respectively) via a maleimide linkage represented by accessible volume (AV) "clouds" are shown (donor: light blue, acceptor: light orange, note with our labelling protocol it is also possible to have donor at A473C and acceptor at A97C; PDB ID: 1UAA. The distance between the mean positions of the accessible volumes (Rmp) are 31.7 Å for closed and 78.5 Å for open. **b.** 10% TBE polyacrylamide gel and plot showing unwinding of radiolabelled forked substrate, to produce ssDNA; wild type and labelled Rep concentrations are 1, 10 and 50 nM. **c.** Unwinding assay using a fluorescence-based timecourse (black wild type, grey double-labelled Rep). **d.** FCS showing normalised autocorrelation, G, plotted as a function of translation diffusion time for donor-labelled 30 nM Rep-AF546 (A97C) from 10-500 mM NaCl (cyan, orange, purple lines) compared against free Cy3B (0 mM NaCl, red) which has a well-defined diffusion coefficient in water, Inset shows the mean (±SEM) diffusion coefficient of Rep- AF546(A97C) (red) and Rep- AF546(A473C) (black) vs NaCl concentration, number of technical replicates n = 3.

We also used FCS, a well-characterised approach that can be applied to probe the mobility of ensembles of solvated biomolecules[51–55], to gain preliminary insights into the effects of changing salt concentration on Rep's structural conformation by measuring its molecular mobility. Although FCS has been used to investigate DNA helicase catalysed unwinding[53] and the interactions between the RecQ helicase, Rep and short DNA substrates[55], it has not been previously applied to Rep dynamics. To exclude any effects due to changes in FRET and increased photophysical instability of the acceptor compared to the donor dye[56], we used labelled Rep comprising just one donor dye at either of the cys-mutated residues 97 or 473 located on subdomains 1B and 2B, and compared these against free dye in solution over low, intermediate and high salt concentrations to determine fluorescence autocorrelation curves (Methods). Since the hydrodynamic radius of diffusing species scales approximately with the cube route of the species mass, FCS is poor at discriminating different types of molecules unless the difference in molecular weight is typically at least a factor of four[57]. With this caveat, we observed a significant difference between Alexa Fluor 546 labelled Rep and free dye (Fig. 1d, using Cy3B because it has a well-defined diffusion coefficient in water), consistent with there being negligible free-dye contamination in the Rep sample and also revealing that the dye brightness showed no consistent variation to salt concentration (equivalent to 16, 11 and 14 kHz/molecule at 10, 150 and 500 mM NaCl, respectively). Fitting of the normalised autocorrelation functions suggested a marginal NaCl-dependent increase in diffusion coefficient and corresponding decrease in hydrodynamic radius of 10-20% depending upon which cys-mutated residue site was used, across the range in salt concentrations tried (Fig. 1d inset). However, since we also found evidence for Rep aggregation in the FCS samples which increased with decreasing salt concentration, these hints at conformational variability of Rep with changing salt concentration warranted a definitive single-molecule approach that could abrogate the issues associated with sample aggregation.

**Single-molecule FRET reveals Rep exhibits four states**

Although FCS is capable of detecting single-molecule fluorescence bursts, both it and ensemble FRET spectroscopy report on population properties that potentially mask molecular heterogeneity[58–61], such as that due to sample aggregation effects. To investigate the dynamics of individual Rep molecules, we first tried dual-colour TIRF and variable-angle fluorescence microscopy[62] to visualise fields of view of labelled Rep molecules immobilised to a coverslip surface (Methods). However, although donor- and acceptor-only excitation generated similar numbers of fluorescent foci per field of view, the yield of detectable anti-correlated behaviour from colocalised donor/acceptor foci was poor due most likely to surface-induced dye photoblinking (Supplementary Fig. 4) as well as proteins labelled with either two donors or two acceptors.

To mitigate against surface effects, we then tried smFRET detection using confocal laser excitation approximately 20 µm deep into the solution on freely diffusing Rep (Fig. 2a,b and Methods). As for the earlier FCS measurements, diffusion of labelled Rep molecules could be monitored as fluorescence emission during the typically ~millisecond transit time through a confocal laser excitation volume. However, here we used rapid µs timescale alternating laser excitation (ALEX) for excitation of the donor and acceptor separately that could be reconciled into coincident millisecond timescale bursts of fluorescence in donor and acceptor channels (Supplementary Fig. 5).

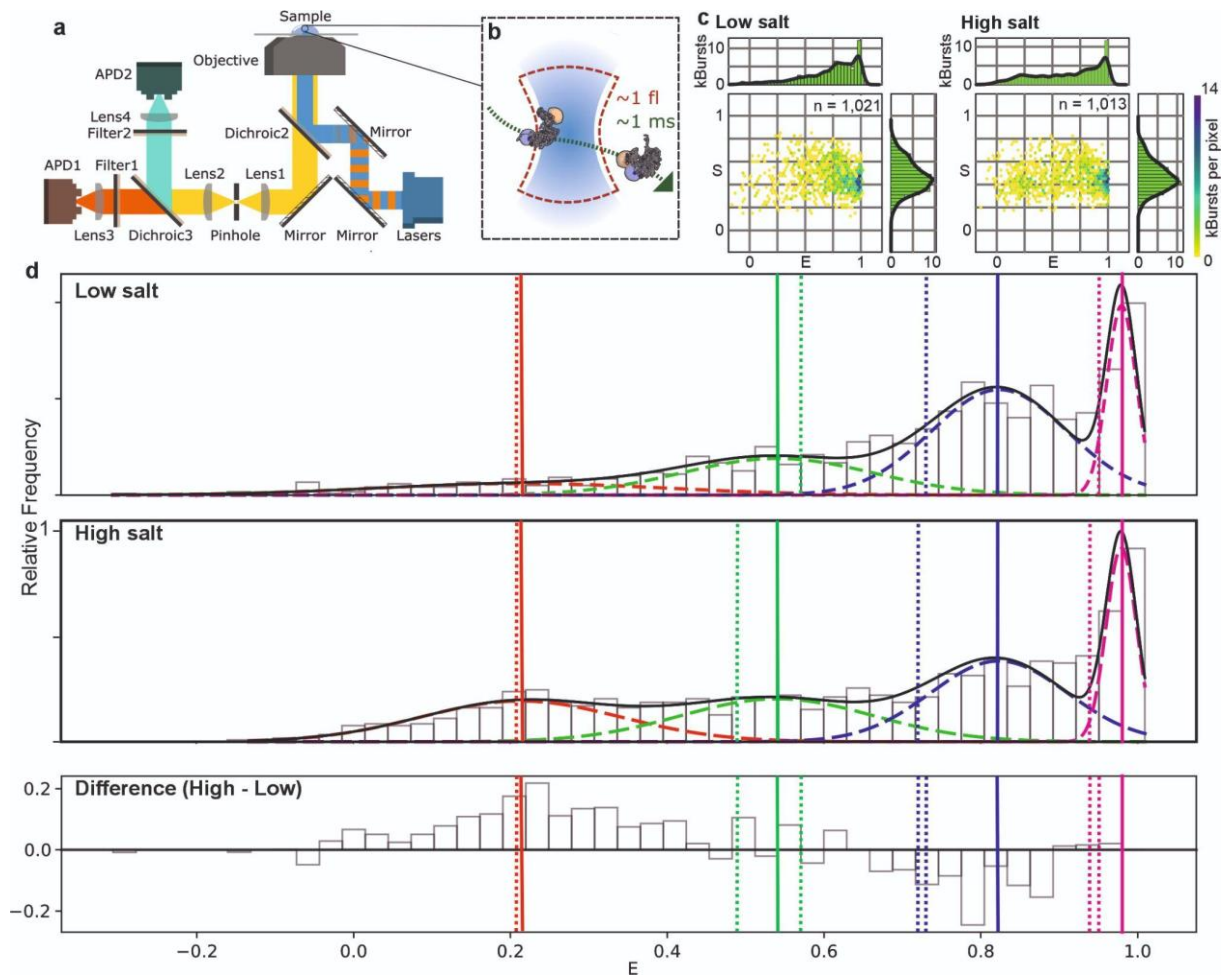

**Figure 2. Single-molecule FRET using confocal microscopy reveal four salt-dependent Rep FRET states. a.** Schematic of the smFRET instrumentation with **b.** a labelled Rep (not to scale) diffusing through the confocal volume. **c**. E-S plots for Rep in low salt 10 mM (number of detected burst events n=12,511) and high salt 500 mM NaCl (n=12,607). Note that the number of burst events is higher than those used in the 1D histograms as they include the singly and doubly labelled populations. **d.** Distribution of FRET efficiency for donor and acceptor labelled Rep (DA-Rep) population using confocal ALEX smFRET in (top panel) low and (middle panel) high salt. A four component Gaussian fit model is overlaid using centre values constrained by global fit parameters (Methods). Dashed lines indicate individual Gaussians for states S1 (red), S2 (green), S3 (blue), S4 (magenta), summed total shown in solid black. Bottom panel shows the difference between low and high salt distributions. Centre of Gaussians indicated as solid vertical lines, dotted lines represent FRET efficiency populations on the individual low and high salt datasets recovered from H2MM with accurate FRET correction parameters applied (Methods); note the high and low salt H2MM S1 FRET predictions were identical to within error hence are overlaid on the same plot. Taken using numbers of bursts n = 1,021 and 1,013 from high and low salt respectively, from 7 different samples.

Photon counts from each channel were converted into estimates of FRET efficiency E and relative stoichiometry S of donor and acceptor labelling on a molecule-by-molecule basis (Fig. 2c). We measured the proportion of doubly-labelled Rep containing both a single donor and a single acceptor dye molecule as just 8% of the total of all labelled Rep molecules, illustrating the value of using ALEX in conjunction with the E-S relation to pull out the specific DA-Rep population for FRET analysis, including filtering out unwanted effects from aggregation and photoblinking. Analysing this DA-Rep population in low salt indicated a broad distribution of FRET efficiency values across the full range of 0-1, but with a bias towards high values (Fig. 2d, upper panel). In high salt, the FRET efficiency distribution was similarly broad but with a bias towards lower values (Fig. 2d, lower panel). Using a multiple Gaussian fit model, we could account for these two distributions with four FRET states S1, S2, S3, S4 of mean values 0.21, 0.54, 0.82 and 0.98 with sigma widths in the range 0.02-0.18 (Supplementary Table 3). The Bayesian information criterion (BIC) value for the global fit to the combined low and high salt data of -234.59 compared to -82.98 and -51.77 for corresponding 3- and 2- component Gaussian fits respectively (Supplementary Table 3). Trying a fifth Gaussian component in the model resulted in a less negative BIC value of -219.34 as well as converging on one of the existing four components; we measured similar trends using other standard statistical metrics of goodness of fit including $R^2$ (approximately 0.9 for the 4-component fits compared to 0.6-0.7 for 2- and 3- component fits) and reduced Chi squared (1-4 for the 4-component fits compared to 4-8 for the 2- and 3- component fits). We therefore concluded that there was reasonable evidence to support a four-component model. The proportions of Rep in each state as assessed by the area underneath each respective Gaussian curve were 11:23:47:18% and 25:25:33:17% (to the nearest %) at low and high salt respectively corresponding to S1:S2:S3:S4. Analysing the differences between the low and high salt data distributions indicated that largest discrepancies were centred around the low FRET state of 0.2-0.3 and the higher FRET state of 0.8-0.9.

We then tried an intermediate NaCl concentration of 150 mM which approximated the physiological saline ionic strength and used this salt level to investigate the effect of addition of DNA substrate. The distribution of E values both with and without the DNA substrate present indicated good fits to the four-component Gaussian model (Fig. 3) with the optimised Gaussian centre values matching those of the low and high salt datasets to within the observed respective sigma widths (Supplementary Table 3). We observed that addition of DNA resulted in a significant decrease in the proportion of smFRET values associated with S1 from 18% to 14% (p = 0.05, permutation test) with a significant increase for S2 from 16% to 22% (p = 0.03), with a less significant decrease in S4 (p = 0.14) and no significant changes with S3 (p = 0.71), however the overall distribution of weights was significantly different (p = 0.05).

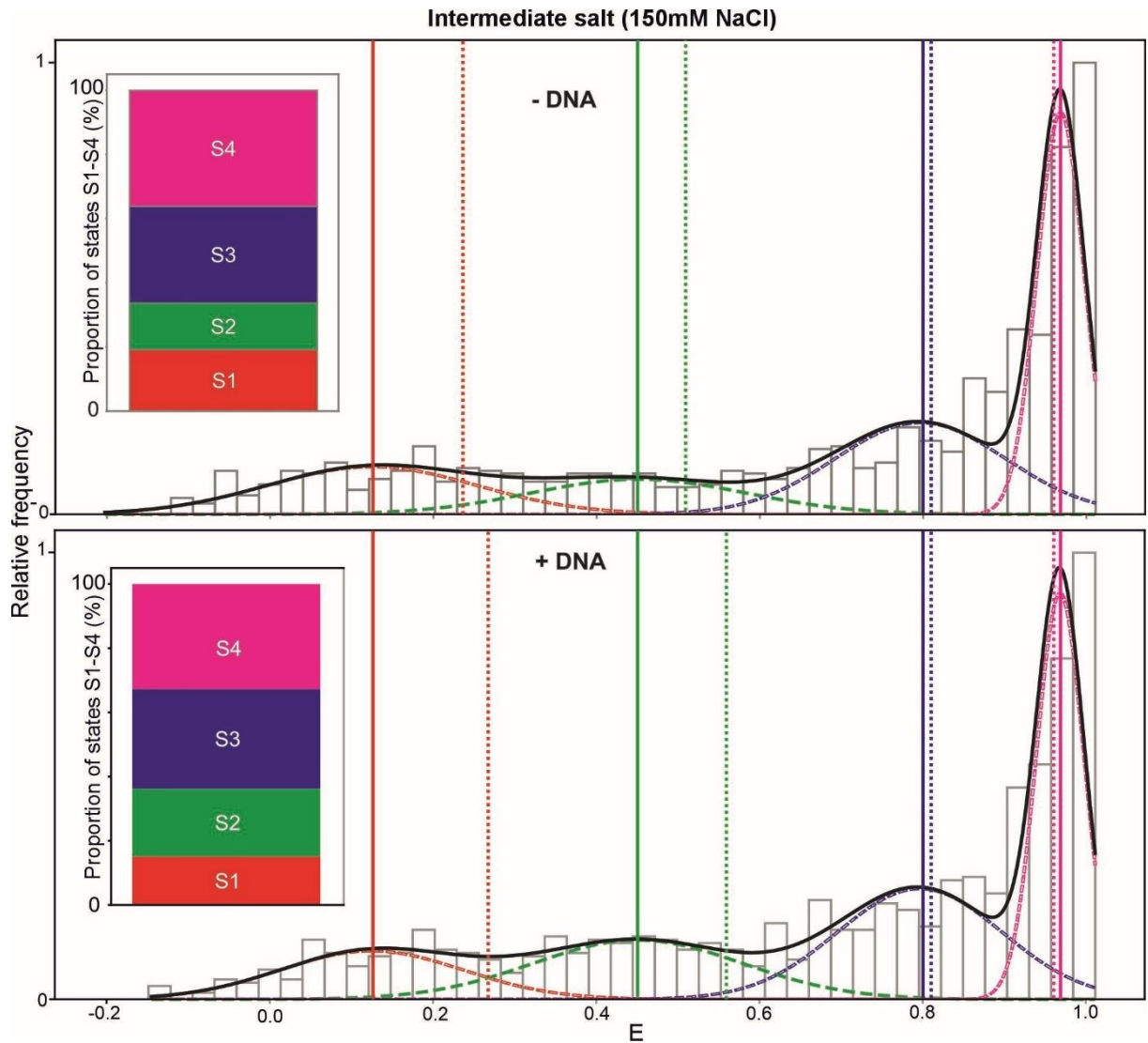

**Figure 3. Addition of DNA substrate results in significant changes to the populations in Rep states S1 and S2.** Distribution of in-solution smFRET efficiency data (E) obtained at an NaCl concentration of 150 mM (grey histogram) can be fitted with to a four-component Gaussian model (red, green, blue magenta dashed lines, total summed Gaussian fit solid black line; Gaussian centre values indicated by solid vertical lines, dotted lines indicate H2MM predictions, see Methods). Inset shows the proportion of E data as a normalised % of the total associated with each state corresponding to the area underneath each respective Gaussian curve.

In order to investigate the lifetime of Rep conformations beyond the typical millisecond timescale transit time through a confocal volume, we were able to monitor single free molecules for extended time durations up to three orders of magnitude greater by using an ABEL trap[63] even for high salt conditions[64], optimising preliminary ABEL measurements on Rep made in low salt non-physiological buffers[45]. The ABEL trap maintains the position of a trapped single molecule in a confocal volume in free solution within a microfluidics cell, away from the perturbative effects of surfaces, and the trapped molecule is able to rotate freely similar to confocal smFRET, but using advanced closed-loop electrokinetic feedback to dynamically recentre a diffusing tracked molecule between four microfabricated electrodes which thus extends the data acquisition timescale compared to standard confocal microscopy. The trap location is conjugate to sensitive detectors that can record simultaneously sample brightness, emission colour, lifetime, and anisotropy, among other spectroscopic and physical variables[65–68]. A diagram of the ABEL trap and examples of raw data from individual trapped DA-Rep molecules are shown in Fig. 4a,b. In the raw trapping traces, anti-correlated brightness of donor and acceptor are evident, along with transitions among multiple FRET states. Although there are differences in the absolute magnitude of apparent FRET values observed due to the excitation laser sources and optical filters used not being identical to those of the in-solution confocal smFRET experiments, these data show that both high and low FRET states are accessed in both salt conditions, but that extended high and low FRET states are more common in low and high salt, respectively.

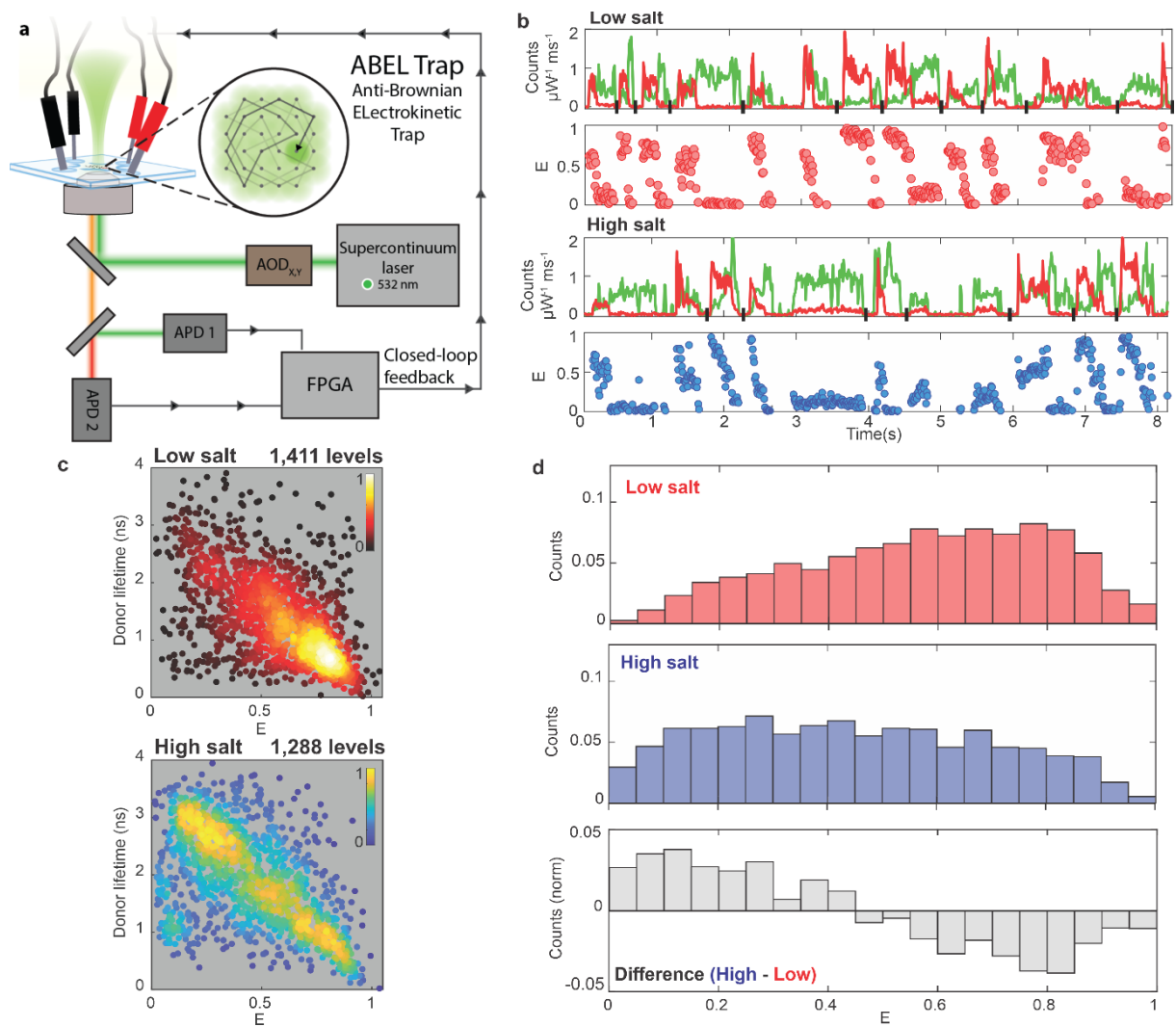

**Figure 4. Lifetime and FRET distributions of Rep measured in the ABEL trap are salt-dependent. a** Diagram of ABEL trap setup as used in this work. Pulsed, linearly polarised 532 nm wavelength excitation is scanned by AODs across a microfluidic trapping region; photon-by-photon readout is used to determine closed-loop voltage feedback to X and Y electrodes to counteract Brownian motion. **b.** Raw ABEL trap brightness data (top) and FRET values (bottom) for several sample events (separated by black vertical slashes on abscissa) show transitions occurring on broad millisecond to second timescales in both low salt, here using 25 mM NaCl (top), and high salt 500 mM NaCl (bottom). Multiple distinct levels are observed, along with dye photophysics. Brightness traces are normalised by excitation power and bin size and corrected for different detection efficiency due to detection filters in each channel. **c.** Scatter plot of measured fluorescence lifetime vs FRET efficiency for DA-Rep population using 532 nm wavelength excitation in low (red; top) and high salt (blue; bottom); each point represents one brightness level; colouration reflects local relative density. **d.** Histogram of FRET

efficiency in low (red) and high (blue) salt, with difference in normalised FRET efficiency histograms for high minus low salt data indicated (grey).

Fig. 4c shows scatter plots of simultaneously measured fluorescence lifetime and FRET efficiency, where each point represents a single level as identified by change point analysis (see Methods). Each scatter plot contains data from >1,000 levels where both donor and acceptor were photoactive. The colour of each point reports the relative local density of points. These data confirm a correlated shift from high FRET efficiency and short donor lifetimes in low salt to a lower, broader FRET efficiency distribution in high salt. One-dimensional histograms showing the FRET efficiencies from this ABEL trap data are shown in Fig. 4d, with equivalent difference histograms, which show a similar qualitative pattern to the result obtained by confocal microscopy with ALEX as described above, but demonstrating that some FRET states can persist for an extended timescale of ~seconds.

**Rep undergoes dynamic transitions over a timescale range of sub-milliseconds to seconds**

Burst Variance Analysis[69] (BVA) on the confocal in-solution smFRET data indicated that the variance of the fluorescence intensity during diffusional transit was broader than theoretical expectations[69] if just a single FRET state was present (Supplementary Fig. 6), suggesting transitions between FRET states during a millisecond timescale. To investigate this more quantitatively, we applied H2MM. The H2MM algorithm is a specialised hidden Markov model designed for analysing confocal single-molecule fluorescence data[70]. It enables characterisation of biomolecular conformations and kinetics by processing the sequential photon-by-photon data from within bursts utilising photon arrival times to determine the dynamic states and transitions of molecules through maximum likelihood estimation[67] (Methods). We used these predictions to infer the likely number of states present and to determine optimised values for the associated kinetic rate constants, based on a modified Bayesian information criterion (BIC')[71] which was checked against Integrated Complete Likelihood (ICL) (Supplementary Fig. 7). Results from H2MM low, intermediate and high salt data indicated independently from the Gaussian fit model that four or five FRET states best described the data (Supplementary Table 4). For some of the datasets it was not possible to probabilistically discriminate between four and five states, however, the lowest FRET state was relatively sparsely populated, and likely represented an acceptor-dark state which was lower than expectations from the mean FRET value for the open crystal structure based on AV estimations and therefore likely to be of less biological relevance; it was therefore excluded from subsequent analysis. Using the four remaining H2MM states (Fig. 5) indicated that two states (consistent with states S1 and S4 from the earlier four-component Gaussian model to within respective sigma widths) exhibited FRET values which were close to expectations for the known crystal structures for closed and open conformations of Rep respectively: the two predicted FRET efficiencies from the crystal structure are 0.29 and 0.96 for the open and closed structures respectively, while the efficiencies from our in-solution smFRET data at 150 mM NaCl including the presence of DNA are 0.27 and 0.96 for states S1 and S4 respectively (see Supplementary Table 4). Fig. 5a indicates the estimated rate constants between all four states from this analysis and the relative occupancy of each state, with the right panel indicating the difference in these constants.

This analysis indicated that the kinetics between S1 and S4 primarily involve transitions between intermediate states S2 and S3, with transitions between S2 and S4 being relatively unlikely. Also, increasing the NaCl concentration results in significant changes to the S1-S3 transitions (both forward and reverse rates are significantly *decreased* in high salt) and the S2-S3 transitions (both the forward and reverse rates are significantly *increased* in high salt). At the intermediate 150 mM NaCl concentration, the S2-S4 and S1-S3 transitions were absent. Interestingly, both the forward and reverse S2-S3 and S3-S4 transition rate constants were higher compared to low and high salt conditions. A key finding from the H2MM on the physiological 150 mM salt concentration (Fig. 5b) is that there is negligible flux between pathways S1→S4, S1→S3 and S2→S4. This implies that, from a starting state of the open conformation (S1), the transitional pathway to the closed conformation (S4) must go via the intermediate states S2 and S3, namely that the full transitional pathway from open to closed involves S1→S2→S3→S4.

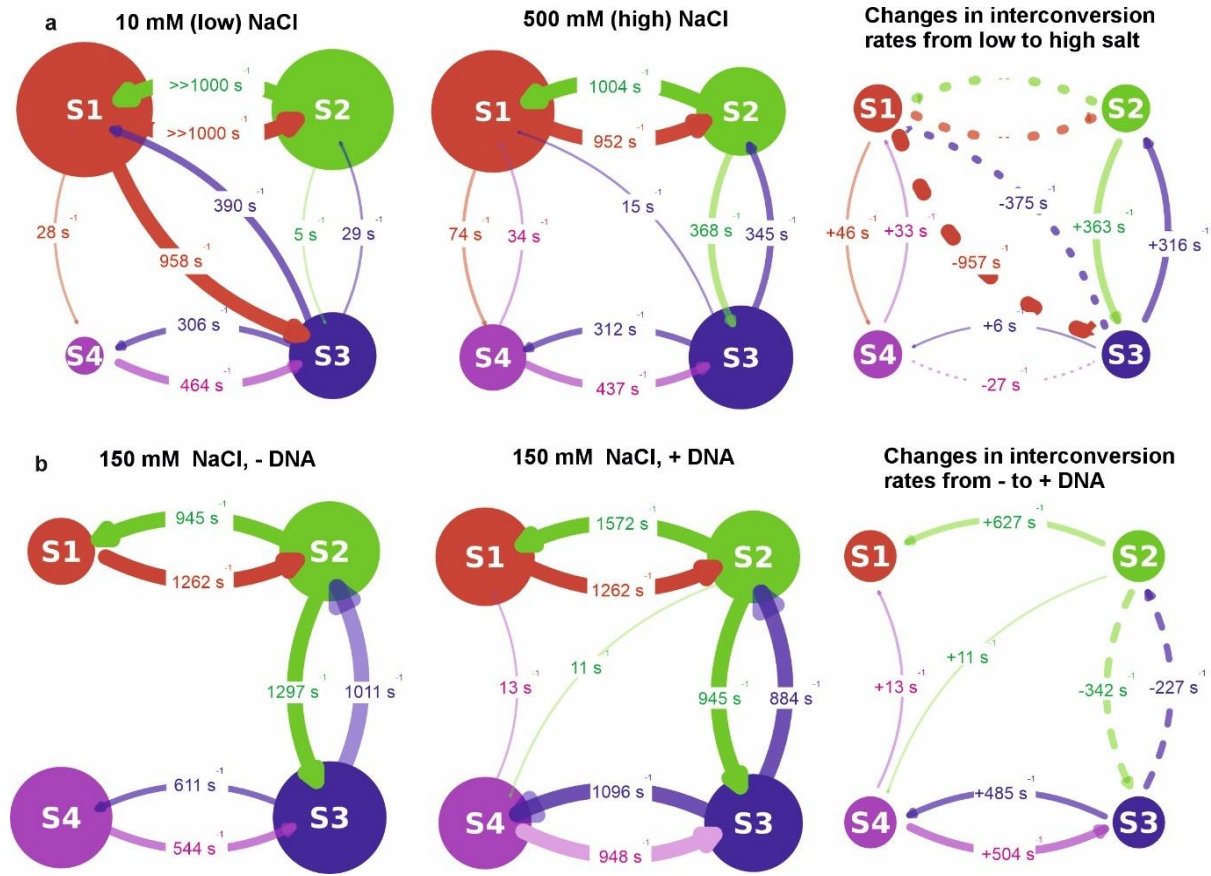

**Figure 5. Rep undergoes sub-milliseconds to seconds timescale FRET transitions between four states**. H2MM on confocal excitation smFRET data at **a.** low (10 mM) and high (500 mM) NaCl concentrations, showing the estimated interconversion rate constants between the four identified FRET states (same colour codes as the states indicated from the previous four-component Gaussian fit model of Fig. 2), with optimised rate constants shown between these (arrow thickness proportional to rate, diameter of each respective state circle proportional to its relative occupancy for the left and middle panels. Right panel shows the change in these interconversion rate constants between low and high salt (solid and dashed arrows are positive and negative differences respectively). Note, since the S1-S2 transitions in low salt typically occur faster than the nominal fastest ALEX sampling time of 100 μs we cannot quantify their absolute change in going from a low to a high salt environment. Similarly, since the S2-S4 transitions are relatively rare at <1 s$^{-1}$, the absolute changes in interconversion rates due to changes in NaCl concentration cannot be quantified, and these very low-rate transitions are not depicted in the plots. **b.** Similar H2MM taken for 150 mM NaCl concentration for -DNA (i.e. DNA absent) and +DNA (i.e. DNA present), with the right panel depicting the difference in rate constants between the two. Addition of DNA results in an increased interconversion rate for S2→S1 by ~50% with ~20-30% lower forward and reverse rates for S2→S3 and ~20-30% higher forward and reverse rates for S3→S4.

**Rep's transition from open to closed is more complex than a simple swing-arm movement of the 2B subdomain**

To establish the mechanistic features of the Rep conformational transitions, we used atomistic molecular dynamics (MD) simulations to Å-level resolution (Methods). The starting structures were derived from the crystal structure of Rep (PDB ID 1UAA[18]) with (i) open and (ii) closed conformations, initially deleting the DNA and then subsequently including it. For capturing large conformational transitions under low, intermediate and high salt, we used 50 ns windows in implicit solvent, running four replicates for each salt concentration and starting from either the open or closed crystal structures. Then, time-dependent conformational changes were evaluated (examples shown in Supplementary Movies 1-6). Note that since these simulations had to run in implicit solvent for computational tractability, the 50 ns timescale does not map to an experimental 50 ns timescale which may be longer by several orders of magnitude because of aqueous solvent viscosity. To make direct comparisons of timescale would require extensive explicit solvent all-atom simulations which extend up to the millisecond experimental timescale, which has an enormous computational burden for even the most powerful supercomputing clusters, and which was beyond the scope of this current study.

A further point to note is that at the high NaCl concentration of 500 mM although there is an anticipated weakening of the electrostatic interaction (through electrostatic shielding manifest as a reduction of the effective Debye length), our MD simulations capture this physical effect, as the open conformation of Rep is more probable when starting with the closed conformation in high salt than in low salt. However, the closed state is not completely inhibited in high salt, which agrees with the fact that DNA can still bind at these NaCl concentrations.

From these simulations we evaluated the straight-line displacement between residues 97 and 473 to test for evidence of any transient intermediate conformations between the labelling sites chosen for smFRET measurements. For simulations which started from the open conformation, we observed significant variability in this displacement within individual 50 ns simulations windows of approximately 5 to 11 nm regardless of NaCl concentration; this was broadly consistent with theoretical expectations for Rep maintaining an open structure (Supplementary Fig. 8). When the simulations started from the closed conformation, the dye displacement was indicative of theoretical expectations of the closed conformation at approximately 3 nm for all of the low-salt replicates. In some of the replicates there was evidence of an increase from 3.5 nm to 8 nm from the closed structure between different datasets of increasing NaCl concentration; this was consistent with theoretical expectations for transitioning to the open structure though interestingly this was a gradual as opposed to a sudden increase which might be anticipated from a distinct "switch" type event between two states; i.e. it is more consistent with the presence of metastable levels between these two quasi-stable states of open and closed.

Estimating the radius of gyration from the MD simulation structures indicated a distribution between approximately 25-45 nm across the different simulation conditions, broadly comparable to the experimental measurements from FCS (Supplementary Fig. 9a-d and Methods), whose distribution could be fitted with up to four Gaussians. This in itself was not direct supporting evidence for the four different FRET states observed experimentally for in-solution smFRET data, however, it was consistent with the presence of a heterogeneity of molecular conformational states[72], and again also a qualitative indicator that the system is more complex than two relatively stable conformations of open and closed.

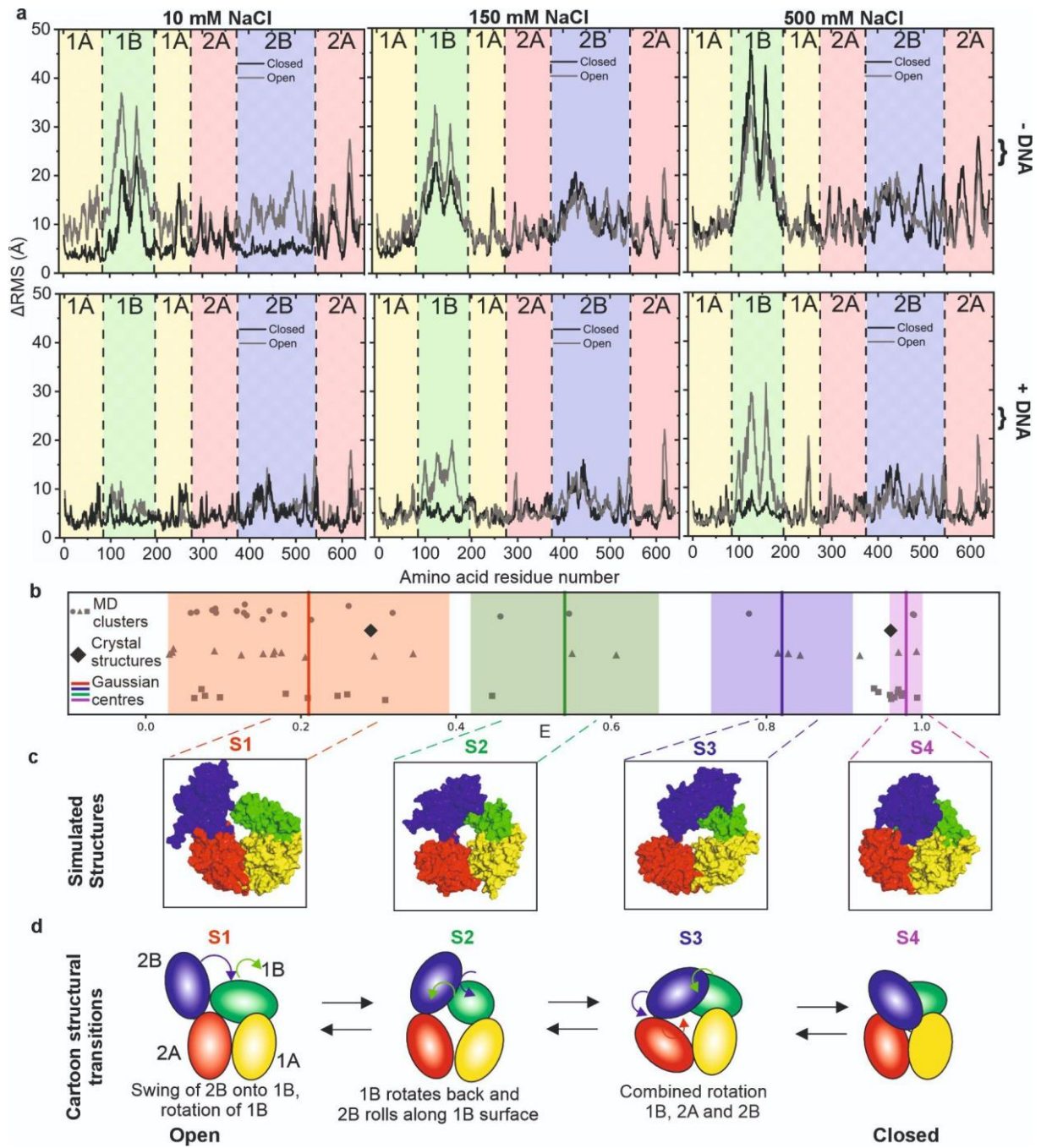

**Figure 6. MD simulations on Rep indicate that the transition between open and closed structures does not involve a simple swing-arm mechanism. a.** The ΔRMSD - root mean squared deviation per residue - averaged over all replicates starting from the open and closed crystal structures suggests significant flexibility in the 1B domain in low (10 mM), intermediate (150 mM) and high (500 mM) NaCl concentrations with no DNA bound to Rep (upper panel) and DNA present (lower panel). Binding of DNA to Rep results in a significant overall reduction in ΔRMSD apart from the 1B domain in the open Rep conformation at high salt. **b.** Cluster analysis on MDS predicted structures followed by FRET AV analysis compared to the experimental observations of four different Rep FRET states, centre of Gaussians from Fig. 2 overlaid in vertical solid lines, shaded areas are standard deviation widths. **c.** Simulated cluster structure whose theoretical FRET efficiency most closely agrees with the mean of each respective experimental FRET state from the four-component Gaussian fit model. **d.** Cartoon schematic depiction of the four Rep subdomains based on the structures selected from panel **c** using animation software to interpolate the conformational transitions between each of the four structures.

Fluctuation analysis using ΔRMSD (root mean square deviation taken for each individual residue average across all replicates, see Methods) of all collected MD trajectories (Fig. 6a) indicated high flexibility in the 1B domain for low, intermediate and high salt conditions in the absence of any bound DNA. In high salt, the open and closed forms of Rep display striking similarities, but at lower salt we saw a decrease in flexibility of the 1B and 2B domain in the closed conformation. The high flexibility of the 2B domain could also be seen from population radar plots from these simulations of the rotation angle of the hinge region of Rep which connect the 2A and 2B domains[17–20,22,23,35,36] (Supplementary Fig. 9e,f and Methods). We note from the crystal structures the presence of an electrostatically charged interface between an alpha helix in the 1A subdomain (position R195-K207) and an alpha helix in the 1B subdomain (position L87- L99) which could therefore be a possible site of action for NaCl concentration sensitivity in rotation around the hinge region. Apart from the high 1B flexibility in the open conformation also being seen at high salt when DNA is bound, DNA binding resulted in a substantial reduction of flexibility for all other domains at both low, intermediate and high salt as well as for the 1B domain in the closed conformation.

We performed cluster analysis to identify the most likely occurring structures from the full suite of simulated data at low, intermediate and high NaCl concentrations, then used FRET AV analysis[73] on these cluster structures to predict the measured values for FRET efficiency using the known positions of the dyes and incorporating realistic flexibility from the linkers used (Methods). We compared these predicted FRET efficiency values with the mean and standard deviation spread for each respective Gaussian curve from the four-component Gaussian fit model analysis (Fig. 6b). This indicated that, from the 60 structure clusters identified (20 for each NaCl concentration), 57 (i.e. 95%) lay within 1 sigma of its nearest identified FRET state, as did the two FRET efficiency values calculated for the open and closed crystal structures, which offered a level of quantitative confidence to the hypothesis that the majority of simulated Rep structures could be explained by a 4-state structural model. Extracting the four individual structure clusters (Fig. 6c) which showed the closest match to predicted FRET efficiency to the mean of each of the four respective Gaussian curves enabled insights into the specific structural transitions involved between each FRET state We used UCSF ChimeraX 's (version 1.10.1)[74] morph function to generate predictions for the structural changes for the four Rep subdomains which result in an increase in FRET efficiency as Rep transitions from the open to closed state ,we then used Blender[75] to animate these transitions using the molecular nodes plugin[76] (Supplementary Movie 7). This indicates a likely transitional pathway from open to closed from states S1→S2→S3→S4, which involves several discrete structural changes as opposed to a single swing-arm rotation of the 2B domain as had been previously suggested for Rep[18]. Depicted as a 2D cartoon schematic (Fig. 6d), the conformational change from the open structure (state S1) involves an initial truncated swing of 2B onto 1B, followed 2B rolling across the 1B surface (state S2); then followed by a combined rotation of 1B, 2A and 2B about their respective long subdomain axes (state S3) prior to ending with the fully closed structure (state S4). The structural geometric interpolations also show that in the initial step of subdomain 2B moving towards 1B, 1B appears to act as a steric impairment to further swing motion of 2B, and so in effect serves to minimise the chances of premature full closure of Rep in the absence of the DNA substrate.

**Discussion**

Here, we investigated the structural dynamics of single molecules of the bacterial DNA helicase Rep. Our initial ensemble level fluorescence lifetime measurements of doubly labelled DA Rep were consistent with the presence of population level structural heterogeneity of the protein. Using low and high NaCl concentrations we were able to tune the fluorescence lifetimes of donor dyes on the protein, donor lifetimes more than doubling under high NaCl in agreement with similar experiments performed previously on the homologous bacterial DNA helicase UvrD which were consistent with suggested conformational changes primarily involving UvrD's 2B subdomain[23]. Similarly, our FCS measurements performed on donor-only Rep to avoid the confounding effects from FRET and increased photophysical instabilities of the acceptor dye showed hints at possible changes in Rep's molecular conformation with changes in NaCl concentration. Future experiments with improved instrumentation to support lifetime filtered FCS might in principle enable FCS-based measurement of anti-correlation behaviour in the donor and acceptor. However, to circumvent the issues with unavoidable aggregation in the Rep sample we decided to focus on the application of single-molecule precise approaches to investigate its structural dynamics further.

While ensemble lifetime measurements provided population-averaged information, our subsequent single-molecule experimental and computational analyses directly resolved the underlying conformational subpopulations and confirmed the salt-dependent conformational landscape suggested from these bulk trends. We first attempted to investigate the different Rep conformations at a single-molecule level using TIRF microscopy, in which Rep was fixed to a microscope coverslip. TIRF microscopy proved useful in that it substantiated our premise that free Rep was conformationally dynamic involving more than one structural state. However, these data were dominated by FRET traces that either were indicative of large aggregates of Rep exhibiting exponential-like decays, or of traces that contained non-correlated FRET fluctuations likely caused by well-established issues of dye photoblinking in the proximity of surfaces such as the glass coverslip[77], which motivated our subsequent attempts to visualise Rep conformational dynamics without having to immobilise the protein onto a surface.

Using confocal in-solution smFRET removed confounding effects of surface immobilisation allowing us to identify four principle states spanning mean FRET values in the range 0-1 that Rep occupies in the absence of its DNA substrate, as opposed to just two states of open and closed suggested from prior crystallographic data[18]. The relative occupancy of these states exhibited sensitivity to NaCl concentration, with a bias towards open or closed depending on if the NaCl concentration was high or low. Using ABEL trapping we were able to extend the observation window for FRET to find evidence for dynamic transitions between different FRET states over a range of timescales up to ~seconds. Using H2MM on the in-solution smFRET data further enabled quantitative estimation of the interconversion dynamics between all four FRET states. We found that dynamically transitioning between the states of the free Rep protein resulted in a bias towards either fully open or closed conformations depending on NaCl concentration, however, the proportion of Rep in open:closed S1:S4 states at low and high salt is not 0:100% and 100:0% respectively, but closer to approximately 11:23:47:18% and 25:25:33:17% (to the nearest %) for S1:S2:S3:S4 (taken as the ratio of areas underneath each respective Gaussian fit). This distribution of conformational states could offer an outcome that there are reasonable numbers of free Rep adopting conformations which are either close to the fully open or closed DNA-bound conformations at any point in time thereby facilitating responsive and rapid adaptation between these two states if required physiologically. For Rep, several transitions, including S1 ↔ S2 and S1 ↔ S3, show more than a 1,000-fold reduction in state interconversion rate at high NaCl concentration, indicating strong electrostatic contributions to conformational exchange. The closest previously published data that can be compared against our single-molecule findings here is not from Rep itself but from the homologous UvrD DNA helicase[23]. In this earlier UvrD study the authors used smFRET measurements from dye pairs positioned at equivalent locations[23] but used TIRF microscopy in conjunction with a much slower sampling time of 32 ms compared to that of our detection system which was 2-3 orders of magnitude smaller. With the caveat that UvrD is homologous, though not identical, to Rep, there is value in discussing some comparisons about the interconversion rate constants between the four states that were identified in the earlier UvrD study[23] at approximately equivalent values of FRET efficiency to within one sigma error. We summarise these key comparisons in Supplementary Table 5. Under these conditions, UvrD appeared to maintain a relatively balanced forward and minimal reverse rates under 60 mM NaCl conditions. For Rep, unlike in the earlier UvrD data, we were able to observe interconversion rate constant trends with respect to NaCl concentrations, for example, the forward and reverse transition for S2→S3 and S3→S4 exhibit a local rate maximum at 150 mM NaCl. This is suggestive of a "sweet spot" in NaCl concentration where the energy barrier between S2 and S3 and (to a lesser extent) between S3 and S4 is minimised allowing rapid transitions between the states, likely due to a comparable contribution from electrostatic and hydrophobic interactions. We also detect low but measurable rate constants for the S1→S4 forward and reverse transitions, while the previous UvrD study did not report any. Also, we observed a large change in interconversion rate constant from intermediate states S2 to S3 with a factor x74 increase with increasing NaCl.

Within the major caveat that Rep and UvrD are not identical proteins, these comparisons suggest that one explanation for these differences might be that our faster sampling reveals the more transient and salt-dependent transitions including both the two intermediate states S2 and S3 and the quasi-static states S1 and S4, which the slower sampling system used for the previous UvrD experiments could not achieve. This premise is to some extent reflected in the range of interconversion rate constants for the earlier UvrD study for the active transitional pathways detected being 0.16-0.85 s$^{-1}$, whereas the H2MM predictions on our Rep data span a range which is over three orders of magnitude higher. However, to definitively test this hypothesis would require a proper like-by-like comparison between Rep and UvrD using comparable analytical approaches in future work. In the absence of MD simulations, it would be easy to have assumed that all of the FRET variations were solely as a result of the 2B subdomain rotating around the hinge, discounting the movement of the 1B subdomain that is clearly evident in simulations. A logical theme therefore for future work will be to perform additional FRET experiments in which dyes are placed in suitable positions to pick up the 1B movement, in addition to the motions of the other subdomains predicted by MD.

We investigated the effect of adding DNA substrate at an intermediate NaCl concentration equivalent to the ionic strength of physiological saline, using an 11 base T-repeat DNA oligo long enough to bind just a single Rep molecule. Addition of DNA resulted in a similar four-state distribution of FRET values. Comparing the relative occupancy between the four states suggested some qualitative similarities to those observed previously for UvrD[23], such as an increase in the occupancy of state S2. However, we also observed a significant decrease in the occupancy of state S1 which was not observed in the earlier UvrD experiments. With the caveat of not comparing identical proteins, such a decrease in S1 occupancy upon addition of DNA might be potentially expected if open Rep is sequestering DNA and bringing it into the binding pocket, closing the conformation. Earlier investigations on UvrD speculated that the 2B subdomain can occupy four distinct conformations, two of which appeared to map to the open and closed crystal structures of UvrD while the others appeared to be metastable intermediates; the fact that none of these earlier findings on UvrD[19,23] reported the relative decrease in S1 occupancy that we observed in Rep [19,23]might hint at a differing mechanism between superfamily 1A helicases and their sequestering of DNA. However, it should also be noted there was also an absence of any investigation of state interconversion kinetics for the previous study on UvrD in the presence of DNA. But again, to properly investigate these apparent differences between Rep and UvrD requires a like-by-like comparison of the two proteins using comparable investigative methods in future work.

In addition to providing evidence for structural heterogeneity in Rep, MD simulated structures when combined with AV modelling allowed useful quantitative correlations to be explored between predictions of likely FRET efficiency values and the underlying structural conformations of the Rep protein. There was clearly stochastic heterogeneity present for the simulated FRET values, however, to have 95% of the consensus cluster of structures simulated in MD come within one sigma width of the corresponding experimental FRET values across all four different states demonstrated a reasonable level of agreement at least to within the constraints of experimental error. Importantly, this enabled us to establish what likely structural changes for Rep could be involved in transitioning from the open state S1 to the closed state S4 via the two intermediate states of S2 and S3. As articulated by others previously[18,28,32,33,78], if one looks at just the open and crystal structures then the most parsimonious structural explanation is that the 2B subdomain exhibits a single swing-arm motion towards the 1B subdomain to bring about closure. This is not to say that there is no potential for metastable states to occur along the line of that swing; however, traditional structural methods will necessarily only capture the more stable of states. This is borne out by free energy simulations carried out on the homologous UvrD that identified a conformationally tilted state of UvrD that facilitates its strand switching and rezipping activity[79] that was previously undetermined by structural studies. This state aligns to structures generated using UCSF ChimeraX's morph structure between our S1 and S2 states with an RMSD of ~3.3 Å. For our current work on the Rep protein we do not see this state as metastable, however, there could be good reason for this; in the tilted state, UvrD's 2B subdomain is interacting with duplex DNA via its GIG motif, with a ssDNA tail bound at the cleft between 1A and 2A subdomains, whereas our MD simulations contain only ssDNA that does not interact with the 2B subdomain. The GIG motif in Rep is also less conserved in Rep (being EIG) which likely has a large part to play in Rep's lower affinity for dsDNA and less specificity for forked substrates[80,81] meaning that this tilted metastable state might also be absent in Rep even when a partial duplex/forked substrate is present.

Our new analysis, which integrates high precision smFRET with atomically precise simulations, reveals that the most likely transitional pathway between the open and closed states is S1→S2→S3→S4, borne out by the absence of a measurable direct interconversion between S1→S4 and S2→S4 at the 150 mM NaCl concentration comparable to the ionic strength of physiological saline. DNA introduces measurable S2→S4 (~1%) and S4→S1 (~2%) bypasses; however, the forward flux still primarily proceeds via S2→S3. Similar analysis at low salt reflects this likely transitional route: Direct S1→S3 and S1→S4 rates ≤1%; S2→S3 progression is individually slow but accumulates over repeated revisits, providing the main forward path to S3. Increasing the NaCl concentration to 500 mM still reflects this as the most likely transitional route, however, the two semi-independent reversible loops of S1→S2 and S3→S4 are established with S2→S3 flux diminished, and the population flow becomes confined mostly within the two loops and so the progression from open to closed states is overall much lower at this high NaCl concentration.

The actual conformational changes involved for the transitional pathway S1→S2→S3→S4 are likely to involve a more truncated initial swing-arm motion of 2B followed by a rolling on the 1B surface, combined with rotations of 1B, 2A and 2B. Furthermore, our analysis suggests that the presence of the S2 intermediate may act to minimise the likelihood of Rep closure in the absence of the DNA substrate, which could have obvious physiological advantages.

Our MD simulations also served to place a spotlight on the potential for salt tunability of Rep's conformational changes, such as the possible roles of specific electrostatically interacting residues at the 1B-2B interfaces. For example, it may be valuable to mutagenise residues Asp110-Lys384 and Gln 133-Pro493 that contribute to putative hydrogen bonds between the two domains, stabilising S2, or Asn104 and Met426 stabilising S3. Alternatively, one could muatgenise the charged interface between the 1A and 1B subdomain. Any changes inactivity because of these mutations could provide valuable insight into the control of the transitions between states and as such inform how Rep controls its entry into a more active closed conformation preventing toxicity associated with hyperactivity[33]. These interventions could be done both experimentally and using MD simulations, however, to perform comprehensive simulation analysis in these charged interfaces would require we estimate approximately 30 residues in total to mutate, but then one would also need to do the appropriate permutational controls of these which is clearly a major undertaking for a future study[33]

More generally, our work exemplifies an important role in structural analysis of molecular machines through integrating accurate simulations with FRET data, both of which are precise at the level of single molecules, and which have the temporal and spatial precision to report on the dynamic mechanisms of conformational transitions. For the Rep molecular machine, in the absence of both submillisecond smFRET data and atomically precise MD simulations it would be easy to assume that the distribution FRET variations were solely a result of the 2B subdomain rotating around the hinge, discounting the movements of the other subdomains that are clear in simulations. Whilst with slower sampling rates for FRET measurements such as those using traditional TIRF microscopy it would not be possible to appreciate the underlying structural plasticity of the Rep protein. Future work to explore other super family 1A DNA helicases using the integrative high precision investigative pipeline we have established here could prove valuable in determining how general the features such as salt-dependent structural intermediates are across the wider family of enzymes. Similarly, there may be value in future studies to probe the mechanistic role of cofactors such as ATP/ADP, to investigate the hypothesis that multiple DNA helicases may be required to act together to bring about duplex unwinding, and to determine whether the closed states of different DNA helicases are needed to promote interactions or confer activity.

**Materials and Methods**

**Labelling Rep**. *rep* ΔCys A97CA473C mutant was purchased as a gene synthesis product in a pEX-K plasmid backbone from Eurofins MWG in which native cysteines at positions 18, 43, 167, 178 and 617 were mutated to L, S, V, A and A respectively as per Rasnik *et al.*[78] as well as the A97C and A473C mutation. Our Rep dye labelling strategy builds on work originally developed by the group of Taekjip Ha (University of Illinois Urbana-Champaign) who has previously performed extensive functional characterisation to show that these residues were suitable labelling positions (see Myong et al, Nature 2005). In brief, the authors selected sites that were either alanine or serine, for mutation to cysteine that were surface exposed and not conserved within UvrD, PcrA or Rep and not contained within any of the known helicase motifs of superfamily 1A helicases, all of their mutants retained helicase activity *in vivo* and *in vitro* as determined by plaque assays with φX174 phage and ATPase and DNA unwinding assays. Likewise control experiments we carried out indicate that the mutated Rep protein can complement for a loss of Rep function *in vivo* (Supplementary Fig. 1) and the unwinding activity of the labelled proteins *in vitro* (Fig. 1) is comparable to that of wild type protein. In our labelling protocol, Rep ΔCys A97CA473C was cloned from this plasmid into pET21a via restriction sites NdeI and XhoI creating plasmid pJLH103. Rep ΔCys A97CA473C was subcloned from pJLH103 into pET14b using restriction sites NdeI and XhoI producing His-tagged Rep ΔCys A97CA473C, pJLH135. Rep ΔCys A97CA473C was also cloned into pBAD24 (Kan$^r$) plasmid for use in complementation assays. The *rep* ΔCys A97CA473C mutant was excised from pEX-K *rep* ΔCys A97CA473C plasmid via sequential NdeI restriction, Klenow treatment followed by PstI restriction. This fragment was ligated into a pBAD24 plasmid that had been sequentially XmaI restricted, Klenow treated and PstI restricted, generating pJLH121. Complementation of *Δrep ΔuvrD* lethality by pBAD constructs was performed as previously described[11]. Briefly, pBAD plasmids were transformed into N6524 and N6556. Both strains contained plasmid pAM403 required for maintenance of cell viability of *ΔrepΔuvrD* strains on rich media. pAM403 is highly unstable and lost by re-streaking on minimal media whilst maintaining selection of pBAD. Colonies were picked and grown in liquid minimal media to stationary phase before being serially diluted and spotted on LB agar containing kanamycin ± arabinose and plates photographed after 24 hours at 37°C. Assays were performed twice.

His-tagged Rep ΔCys A97CA473C was overexpressed from the relevant pET14b clones in HB222 (BL21AI*Δrep*). For overexpression, cells were grown to OD$_{600}$ 0.7 shaking at 37°C, induction performed at 20°C with 0.2% arabinose for 3 hours. Cells were pelleted, and flash frozen in 50 mM Tris pH 7.5 and 10% (w/v) sucrose. Cells were thawed on ice and additions then made so that the suspension contained 50 mM Tris-Cl pH 8.4, 20 mM EDTA pH 8.0, 150 mM KCl and 0.2 mg ml$^{-1}$ lysozyme. After 10 min incubation on ice, Brij-58 was added to 0.1% (v/v; final concentration) with a further 20 min incubation on ice. Supernatant was recovered by centrifugation (148,000 x g, 4°C for 60 minutes) and DNA precipitated by dropwise addition of polymin P to 0.075% (v/v; final concentration) with stirring (4°C, 10 minutes). The supernatant was recovered by centrifugation (30,000 x g, 20 minutes, 4°C) and solid ammonium sulphate added to 50% saturation with stirring (4°C, 10 minutes). The pellet was collected by centrifugation (30,000 x g, 20 minutes, 4°C) and stored overnight at 4°C. The protein pellet was resuspended in 20 mM Tris-HCl pH 7.9 and 5 mM imidazole until the conductivity matched 20 mM Tris-HCl pH 7.9 and 500 mM NaCl (buffer A) plus 5 mM imidazole, as determined using a conductivity meter. Rep ΔCys A97CA473C was purified by chromatography on a 5 ml His-trap FF column (GE Healthcare) using a 100 ml gradient of 5 mM to 1 M imidazole in buffer A. The conductivity of eluted protein from the His-trap column was adjusted to match buffer B (50 mM Tris pH 7.5 and 1 mM EDTA) plus 50 mM NaCl by dilution in buffer B, as determined using a conductivity meter. Rep proteins were then purified on a 3ml heparin-agarose column using a 60 ml linear gradient of 50 mM to 1M NaCl in buffer B. Peak fractions containing pure protein were collected and reduced for two hours at 4 °C by addition of 5 mM Tris(2-carboxyethyl)phosphine hydrochloride (TCEP). Ammonium sulphate was then added to the Rep solution under stirring to 70% saturation at 4 °C, stirring was allowed to continue for 10 minutes before centrifugation (18,000 x g, 20 mins at room temperature). The pellet was resuspended in degassed labelling buffer (100 mM sodium phosphate pH 7.3, 500 mM sodium chloride, 20% (v/v) glycerol).

We explored several FRET dye combinations initially using confocal smFRET (Supplementary Fig. 2) before settling on Alexa Fluor™ 546 and Alexa Fluor™ 647. The various dye pairs tested impact not only the total count but also the shape of the distributions. These changes may reflect slightly different R0 or linkers/chemistry. However, one can possibly discount differential labelling of lysines since all dyes are linked by the same maleimide chemistry. This optimised combination of FRET dye pair exhibited not only excellent photophysical properties for smFRET but also that showed the smallest propensity for aggregation compared against the other combinations tried; this combination of factors was the primary reason for focusing on just this FRET pair in subsequent fluorescence measurements. For optimised incubation conditions, Alexa Fluor™ 546 $C_5$ Maleimide and Alexa Fluor™ 647 $C_2$ Maleimide were dissolved in anhydrous DMSO and mixed in equimolar amounts; dyes were added to the reduced Rep sample in a fivefold molar excess. The mixture was rocked at room temperature 30 minutes before addition of 2-mercaptoethanol to 10 mM to quench the labelling reaction and left to rock for 10 minutes. The sample was loaded onto a 1 mL His-Trap FF crude column (GE Healthcare) equilibrated with buffer A (20 mM Tris-HCl pH 7.9 and 500 mM NaCl, 20% (v/v) glycerol) + 5 mM imidazole. The column was washed with buffer A + 5 mM imidazole, until such time as there was no longer any absorbance corresponding to free dyes in the buffer exiting the column (~20 mL), before being developed with a 20 mL gradient of buffer A + 5 mM imidazole to buffer A + 500 mM imidazole. Peak fractions containing labelled Rep were pooled aliquoted and stored at -80 °C in 20 mM Tris-HCl pH 7.9 and 500 mM NaCl, 30% (v/v) glycerol. Protein concentrations were measured using Bradford's assay (with BSA used to provide a standard curve) and, in combination with individual dye concentrations as determined by absorbance measurements using NanoDrop™ 2000, labelling efficiencies and stoichiometries could be determined. It is possible that that there were cooperative effects at play that affect the rate of conjugation of the second dye dependant on the binding of the first dye, however, it was beyond the scope of this manuscript to investigate fully.

**Helicase activity assays.** Unwinding of streptavidin-bound forks was assayed using a substrate made by annealing oligonucleotides PM187B20 (5' GTCGGATCCTCTAGACAGC(biodT)CCATGATCACTGGCACTGGTAGAATTCGGC) and PM188B34 (5' AACGTCATAGACGATTACATTGCTACATGGAGC(biodT)GTCTAGAGGATCCGAC). PM187B20 was 5' end labelled using 1 μL T4 polynucleotide Kinase (NEB, M0201S), 1 μL γ $^{32}$P ATP (Perkin Elmer, NEG502Z250UC), 1 μg PM187B20 in a 20 μL reaction volume of 1x T4 PNK buffer (NEB, M0201S) for 60 minutes at 37 °C before heat inactivation at 65°C for 25 minutes. The reaction volume was brought up to 50 μL by addition of MilliQ water and run through a Micro Bio-Spin™ 6 column (Biorad, 7326221) as per manufacturer instructions to purify radiolabelled oligonucleotide from free γ $^{32}$P ATP. For annealing, labelled PM187B20 was mixed with 3 μg of PM188B34 in 100 μL final volume of 1x sodium citrate, heated to 95 °C five minutes before slowly cooling to room temperature. Annealed DNA was purified from a 10% Acrylamide TBE gel. Reactions were performed in final volumes of 10 μL in 50 mM Tris pH 7.5, 10 mM magnesium chloride, 2 mM ATP, 0.2 mg ml$^{-1}$ BSA and 1 nM DNA substrate. The reaction mixture was pre-incubated at 37°C for five minutes, then Rep helicase was added, and incubation continued at 37°C for 10 minutes. Reactions were stopped with 2.5 μl of 2.5% SDS, 200 mM EDTA and 10 mg ml$^{-1}$ of proteinase K and analysed by non-denaturing gel electrophoresis on 10% polyacrylamide gels. Gels were quantified as previously published[33,82].

Fluorescence-based timecourse assays were carried out using a substrate made by annealing the two oligonucleotides PM187-BHQ1 (5' BHQ1-GTCGGATCCTCTAGACAGCTCCATGATCACTGGCACTGGTAGAATTCGGC) and PM188-AF488 (5' AACGTCATAGACGATTACATTGCTACATGGAGCTGTCTAGAGGATCCGAC-AF488), the AF488 dye on PM188-AF488 is quenched when the oligonucleotides are annealed thanks to the proximity of the 5' Black Hole Quencher 1 on PM187-BHQ1. Unwinding was carried out as above but at room temperature in 100 μL reaction volumes using a BMG Labtech Clariostar Plate Reader to monitor the fluorescence emission from the Alexa Flour 488 at 15 second intervals. The process of labelling takes several hours at room temperature; using the assays whose results are indicated in Fig. 1; it is reasonable to conclude that the Rep protein remains stable over this period as labelled Rep retained similar activity to that of wild type Rep that had not been subjected to long periods at room temperature. Assessing the stability of Rep beyond the 1-hour duration of the fluorescence-based activity timecourse may be valuable in future investigations requiring longer timescale information.

**Ensemble Steady-State and Time-Resolved Fluorescence Spectroscopy.** To verify the donor: acceptor ratio and protein concentration, absorption spectra were acquired using a Cary 60 UV-Vis spectrophotometer (Agilent Technologies) over a 200-800 nm wavelength range with 0.5 nm intervals. Rep concentrations were estimated via Beer-Lambert's law using extinction coefficients 112,000 M$^{-1}$cm$^{-1}$ and 270,000 M$^{-1}$cm$^{-1}$ for Alexa546 and Alexa647, respectively. Steady-state and time-resolved fluorescence spectroscopy was performed using a FluoTime300 spectrophotometer equipped with a hybrid PMT detector (PMA Hybrid 07, Picoquant). Fluorescence emission spectra were measured using continuous 532 nm wavelength laser excitation under magic angle conditions. Ensemble FRET efficiencies, E, were estimated as E = A /(D+A), where D and A represent integrated spectral intensities between 550-650 nm and 650-800 nm, corresponding to donor and acceptor emission, respectively. Time-resolved fluorescence decays were recorded using n = 3 technical repeats throughout by time-correlated single-photon counting (TCSPC) using excitation from a picosecond pulsed laser (Picoquant, LDH-D-FA-530L) coupled to the entry port of FluoTime300 spectrophotometer. Samples were excited with 30 MHz repetition rate. Fluorescence emission decays at 572 nm were collected until 10$^4$ counts accumulated at the decay maximum. Fluorescence decay curves were fitted by iterative re-convolution of the instrument response function (IRF) and the observed fluorescence decay assuming a multi-exponential decay function of the form

$$I_t = \sum_{i=1}^{n} \alpha_{i} e^{-\frac{t}{\tau_i}}$$

where $I_t$ is the fluorescence intensity as a function of time, $t$, (normalized to the intensity at $t$=0); $\tau_i$ the fluorescence lifetime of the $i^{th}$ decay component, and $\alpha_i$ is the fractional amplitude of that component. The fit quality was judged based on convergence of the reduced Chi squared. The different lifetime components and their associated amplitudes are provided in Supplementary Table 2.

**Fluorescence correlation spectroscopy (FCS).** FCS measurements on the doubly labelled complex were performed at room temperature in solution of viscosity 0.001 Pa s using a Zeiss LSM 880 microscope equipped with a GaAsP detector. Normalized FCS curves associated with freely diffusing

Alexa546-Alexa647 Rep helicase were recorded using 30 nM solutions in low or high salt buffer (20 mM Tris-HCl pH 7.9) using 514 nm laser excitation wavelength of typical 4 mW power at the sample plane. The confocal volume was measured using a calibration sample of 6 nM Rhodamine-6G at 21°C with diffusion coefficient 400 µm$^2$s$^{-1}$ [83]. FCS measurements on donor-only (Alexa-546 Rep helicases) were performed under identical solution conditions using an EI-FLEX single molecule spectrophotometer (Exciting Instruments) equipped with 520 nm wavelength laser diode (LuxX, Omicron) excitation. Photon arrival times were recorded on an avalanche photodiode (AQRH-14, Excelitas). Correlation curves G($\tau$) were best fitted to two component, 3D normal FCS diffusion models using the pulsed interleaved excitation analysis with MATLAB (PAM) software[84]. All buffer solutions were assumed purely viscous as opposed to viscoelastic[85], and the variation in solution viscosity due to the range of NaCl concentration used was assumed to be negligible[86]. As a point of note, lifetime filtered FCS might in principle facilitate the observation of fluctuations in the DA/AD anti-correlation signatures for Rep, though this was beyond the scope of our current manuscript to investigate.

**Variable-angle fluorescence and TIRF microscopy.** Our strategy for solution-based imaging of single-molecule doubly labelled surface-immobilised Rep was based on site-specific and ionic interaction between His-tagged Rep and an anti penta-His tag antibody coupled to a glass coverslip via biotin-avidin[87]. Biotinylated polyethylene glycol (PEG) non-specifically adsorbed to pre-silanised surfaces binds to the multivalent avidin protein, providing a specific binding site for the biotinylated anti-His tag which ionically attaches to 6X His antibody on Rep's N-terminus, allowing for conformational freedom for Rep. We used a home-built system enabling control of the angle of incidence by lateral displacement of a downstream lens on a micrometer lateral translation mount[88,89] to generate objective lens mediated Variable-angle fluorescence/TIRF with molecular detection precision, based around a "SlimVar" design[62] but using a broader field of illumination. The angle of incidence was typically set at approximately 45° for variable-angle fluorescence microscopy, with depth of penetration of approximately 100 nm for TIRF, excitation intensity at the sample set at 160 W cm$^{-2}$. Lasers used were Obis LS 532 nm and Obis LX 640 nm at nominal 5 mW power. Fluorescence was collected through a Nikon CFI Apochromat TIRF 100XC oil objective lens, dichroic mirror (Chroma, ZT532/640rpc), propagated through a Photometrics Dual-View colour splitter (dichroic mirror cut-off wavelength 640 nm, donor channel emission bandpass filter of 585/65 nm, and acceptor channel 647 nm long pass) and imaged at an exposure time of 50 ms per frame onto a Prime 95B sCMOS camera at magnification 50nm/pixel. Recorded images were analysed in MATLAB (R2019a) using iSMS single-molecule FRET microscopy software[90]. Briefly, emission channels were aligned, and background-corrected donor and acceptor fluorescence trajectories obtained by integration of pixel intensity within a 9 x 9-pixel area centre on the detected fluorescent foci signals for each time point. Apparent FRET efficiencies were then calculated as described previously. Variable-angle fluorescence and TIRF microscopy both used an oxygen scavenging system comprising 833 µg/mL glucose oxidase, 33 µg/mL catalase, 33 mM glucose and 1 mM Trolox (pH 9.5).

**Confocal Microscopy smFRET.** Freely diffusing smFRET microscopy was performed using a home built confocal microscope, and the EI-FLEX as previously described[91]. Beam illumination was stopped down to a radius of 5 mm (compared against the objective lens back aperture of ca. 10 mm) and so it is underfilled which is unlikely to be a cause of data reduction quality compared to other implementations which involving overfilling. Individual photons were time-stamped to 10 ns precision. The ALEX cycle period was set nominally to 100 µs (comprising 45 µs donor excitation, 5 µs lasers off, 45 µs acceptor excitation, 5 µs lasers off). Individual photon bursts were typically ~1 ms long. Labelled Rep was diluted to ~50 pM in low, intermediate of high salt buffer and imaged in 1-hour acquisitions. Multiple hour-long acquisitions were collated (7 and 6 for low and high respectively) to collect >1,000 doubly labelled bursts for each condition. Data were analysed using FRETBursts[92] in Jupyter notebooks. Background was determined by fitting an exponential decay to interphoton delay distributions, ignoring the short delays corresponding to fluorescent molecules to determine the rate of background photons. Background rates were determined as such every 5 minutes. After background determination and subtraction, doubly labelled bursts were identified using a dual channel burst search (minimum 50 photons in each channel) and accurate FRET correction factors were determined[93]. Fully corrected FRET efficiencies were then subjected to Gaussian mixture modelling using the GaussianMixture function from the scikit-learn Python library[94]; global fits were carried out on the high/low NaCl data and on the intermediate NaCl ± DNA data with the BIC value used to determine the best number of fits (4 in each case), checked against standard statistical metrics of $R^2$ goodness of fit and reduced Chi squared. The Gaussian centres of each global fit were then used to constrain the Gaussians in the individual data sets. Weights represent the fraction of total probability assigned to each Gaussian and their distributions analysed using permutation testing[95,96].

**BVA and H2MM of smFRET confocal microscopy data.** To investigate if freely diffusing Rep was exhibiting millisecond timescale dynamics, the BVA[65] was used using MATLAB program PAM[84]. For H2MM, first an all photons burst search (APBS) was performed with the rate threshold for bursts (F) set at 6 times greater than the background count rate, AA photons ≥50, burst size ≥50, and stoichiometry between 0.2 and 0.8. Using the H2MM_C package[97] models were optimised for 1 to $n+1$ states, where $n$ is the lowest number of states to achieve the BIC' < 0.005 threshold, indicating that additional states provide negligible improvement in model likelihood relative to the increased complexity of an additional state. The integrated completed likelihood (ICL) was also computed as a complementary model-selection metric, incorporating both the global likelihood and the most probable photon-state sequence (via the Viterbi path), with the optimal model corresponding to the minimum ICL value. Log-likelihood uncertainty estimation was performed; this systematically varies each fitted parameter while refitting the model and tracking how the log-likelihood changes, identifying the parameter values where it drops by 0.5 relative to the optimum and corresponds to the one-sigma confidence interval. From this we extracted uncertainties for (i) the determined FRET efficiencies, and the symmetric ± uncertainty for each state's FRET efficiency has been added and (ii) the transition rates, and the lower / upper transition-rate limits for all determined rates have also been added. In principle, photon-by-photon H2MM can resolve microsecond transitions (up to $10^5$ $s^{-1}$). This limit comes from data and simulations demonstrating microsecond resolvability at ~$10^6$ detected photons $s^{-1}$, whereas practical limits scale with detected photon rate and the photons-per-dwell needed for state discrimination. In our data the median total count rate is ~$5.0 \times 10^4$ $s^{-1}$ (10th–90th percentile ≈$2.7 \times 10^4$–$1.0 \times 10^5$ $s^{-1}$); taking a conservative requirement of ~10–20 photons per state dwell for robust discrimination across four states gives a $k_{max}$ (the fastest transition rate that can be reliably distinguished between states given the photon statistics of the measurement) ≈$2.5$-$5 \times 10^3$ $s^{-1}$, i.e. a few thousand per second.

**ABEL trapping, sample preparation and analysis.** The ABEL trap design was similar to previous reports[64,68,98]. 532 nm wavelength laser excitation was provided by a pulsed (60 MHz, ~100 ps pulse width) supercontinuum laser (Leukos ROCK 400-4) paired with a computer-controllable acousto-optic tunable filter (AOTF; Leukos TANGO VIS). A pair of acousto-optic deflectors (AODs) (MTT110-B50A1.5-VIS) arranged orthogonally and controlled by a direct digital synthesizer (DDSPA2X-D8b15b-34) were used to scan the focused laser spot at the sample in a ~3x3 µm$^2$ optimized knight's tour pattern[99]. The sample region was aligned confocal to two avalanche photodiodes (Excelitas SPCM-AQRH-14-ND) in two color channels. Emission from each trapped molecule passes through the dichroic in the microscope (Chroma ZT532/640rpc-UF1), gets filtered through a 300 µm diameter pinhole, is split via a dichroic mirror (Chroma T610lpxr) into green and red channels which are further split into parallel and perpendicular emission channels using polarizing beam splitters (Thorlabs PBS251). Filters in each detection channel are as follows: Green Channel: Semrock BLP01-532R-25, Chroma ET542lp, FF01-571/71-25; Red Channel: Chroma ET655lp, Semrock FF01-660/52-25.

Upon detecting a photon, the APDs send a signal to the TCSPC (Picoquant Multiharp 150) and to an FPGA (NI PCIe-78656) for closed-loop feedback. For each photon, a position estimate was determined based on the laser position when the photon was detected and a pre-calibrated lag. A Kalman filter was used to reduce noise in position estimates [99]. X and Y feedback voltages, proportional to distance from the centre, were applied via the FPGA output to a 10x voltage amplifier (Pendulum F10AD) attached to four platinum electrodes placed orthogonally in X and Y reservoirs of the microfluidic cell (in-house fabrication similar to Cohen and Moerner (2005)[100]). Immediately before trapping, labelled protein was diluted in a buffer containing 10 mM tricine (pH 8.0), 1.25 mM $MgCl_2$, 30 µM KCl, 100 µg/ml BSA, glycerol at 5% for high salt experiments and 10% for low salt experiments, an oxygen scavenging system (60 nM protocatechuate-3,4-dioxygenase and 2.6 mM protocatechuic acid) to final concentration of approximately 700 pM, and NaCl added (25 mM for low salt; 500 mM for high salt). To prevent protein from sticking to the microfluidic cells, cells were passivated using mPEG-silane (Laysan Bio MPEG-SIL-5000, 50 mg ml$^{-1}$) in 95% ethanol, 5% water, and 10 mM acetic acid[101], incubated for approximately 48 hours then in 1% Tween for 10 minutes and rinsed with ultrapure water.

Data analysis was performed using custom software written in MATLAB (Mathworks). Photon arrival times recorded by the Multiharp were used to construct binned time traces. We subtracted the background counts from each of the emission channels. Background bins were identified using an information-criteria-optimized (AIC) K-means clustering algorithm. Bins of constant brightness were determined using a changepoint algorithm that identifies the bins where a substantial change in brightness occurs[102]. Bins between two changepoints are considered a level and get averaged to assign a single brightness or $E_{FRET}$ value for the whole level. The following corrections were performed to the calculated $E_{FRET}$: Donor leakage into the acceptor channel was removed using the emission detected in the acceptor channel for donor only molecules. The differences in detection efficiency of the parallel and perpendicular channels were corrected using the *G* factor. We also corrected for differences in transmission of donor and acceptor emission based on the filters used (71 % transmission for AF546; 55 % transmission for AF647). Lifetime delays for all photons in each identified level were collected into a histogram and fit with a single exponential decay lifetime convolved with the measured instrument response function, using a maximum likelihood algorithm after Zander et al.[103] and Brand et al.[104] as previously described[64,66,67]. Calculated parameters from levels longer than 30 ms are plotted as scatter heatmaps or histograms as shown in Fig. 3. Since there are differences in the excitation laser sources and optical filters used compared to those for confocal smFRET the precise values of the apparent FRET efficiency values generated will not be identical. Also, since the confocal smFRET calculations use much shorter bursts compared to ABEL data these have a much larger statistical spread.

**MD Simulations.** Simulations were generated using AMBER 18 [105] with CUDA supported MD engine, pmemd, using the ff14SB and bsc1 forcefields for the protein and DNA parameters, respectively[106,107]. Open and closed Rep starting structures were from PDB ID 1UAA[18]. Each structure had missing amino acids in the loop connecting 2A and 2B subdomains, with the closed structure missing eight (M539, M540, E541, R542, G543, E544, S545, E546) and three in the open (G543, E544, S545). These gaps were filled by incorporating the loop from the closed structure into the open, with the remaining three amino acid gaps in both structures filled by loops from the homologous protein UvRD (PDB ID 2IS2). The protonated state was determined using the Virginia Tech H++ web server[108]. Simulations were carried out on open and closed structures separately with and without DNA at 300 K and 1 atm[109]. The system was solvated implicitly using the generalised Born model with GBneck2 corrections with an infinite cut-off, set with mbondi3 radii. There were two monovalent salt conditions, low (0.01 M) and high (0.5 M). Simulations were minimised and equilibrated using standard protocols[110], run for 50 ns. Temperature was maintained at 300 K with a Langevin thermostat using collision frequency 1.0 ps$^{-1}$. Four replicas were run for each of the eight conditions: the protein with/without DNA starting from open/closed in high/low salt, totalling 32 simulations and 1.6 µs. The Cpptraj program was used to calculate all analysis including RMSD, radius of gyration and frame clustering[111]. The latter was performed via hierarchical agglomerative clustering based on the average linkage algorithm using RMSd between frames as a distance metric. Angles between 2B and 2A were measured using the centre of mass of each subdomain.

**FRET accessible volume (AV) analysis**

AV and FRET efficiency predictions from the MD simulations were calculated using the FRET modelling software Olga[73]. Fluorophores with their respective linkers were implemented onto modelling structures of Rep with the parameters indicated in Supplementary Table 6). The software calculates AVs by modelling the dye as a sphere of given radius where the central atom of the fluorophore is connected by a linker with a given length and width to the attachment atom of the modelled structure, finding all positions where there are no clashes between the dye, molecule surface, or linker. This is repeated with 3 different radii characteristic of the fluorophore. Predicted FRET efficiencies are calculated by taking the average of FRET efficiencies computed from pairwise distances between positions sampled from the two AVs.

**Acknowledgements**

We thank the Bioscience Technology Facility, University of York for technical support.

**Funding**

BBSRC grant BB/P000746/1 (MCL)

Neubauer Family Foundation (AHS)

NSF QLCI QuBBE grant OMA-2121044 (AHS)

EPSRC grants EP/N027639/1, EP/T022205/1, EP/R029407/1, EP/P020259/1 and EP/V034030/1 (AN)

EPSRC grant EP/N509802/1 (AA-B)

DFG - German Research Foundation grant 528591139- FIP 31 (MD)

Carl Zeiss Stiftung grant P2022-13-017 (MD)

BBSRC grants BB/W00061X/1 and BB/T008032/1 (TDC)

EPSRC grant EP/V034804/1 (TDC)

University of York Research Priming funds grants 50109444 (AN)

University of York Centre for Future Health grant 204829 (AN)

Alzheimer's Research UK grant RF2019-A-001 (SDQ)

**Author contributions**

Conceptualization of study: MCL, JALH, SDQ

Conceptualization of MD simulations: AN

Data collection and analysis: JALH, SDQ, LD, BA, MASA, MD, AE, AS

MD simulation data collection and analysis: LF, AA-B

Writing – original draft: JALH, SDQ, ML

Writing – review and editing: all authors

Project administration: MCL

**Competing interests**

All authors declare that they have no competing interests.

**Data and materials availability**

Experimental data used in this article can be accessed from https://doi.org/10.5281/zenodo.12666478. Simulation data are at https://doi.org/10.15124/28e00c09-af29-4dd0-a7b5-c90047f85fbf. All plasmids available upon request subject to restrictions including completion of materials transfer agreements.

# Supplementary Information

**The transitional kinetics between open and closed Rep structures can be tuned by salt via two intermediate states**

Jamieson A L Howard[1], Benjamin Ambrose[2,3], Mahmoud A S Abdelhamid[2], Lewis Frame[1], Antoinette Alevropoulos-Borrill[1], Ayesha Ejaz[4], Lara Dresser[1], Maria Dienerowitz[5], Steven D Quinn[1,6], Allison H Squires[4,7], Agnes Noy[1,6], Timothy D Craggs[2], and Mark C Leake[1,6,8] †

[1] School of Physics, Engineering and Technology, University of York, York, YO10 5DD, UK

[2] Department of Chemistry, University of Sheffield, Sheffield S3 7HF, U.K.

[3] Current address: Single Molecule Imaging Group, MRC-London Institute of Medical Sciences, London, W12 0HS, UK.

[4] Pritzker School of Molecular Engineering, University of Chicago, IL, USA.

[5] SciTec Department, Ernst-Abbe-Hochschule, University of Applied Sciences, Jena, Germany.

[6] York Biomedical Research Institute, University of York, York, YO10 5DD, UK.

[7] Institute for Biophysical Dynamics, University of Chicago, Chicago, IL, USA.

[8] Department of Biology, University of York, York, YO10 5DD, UK.

† For correspondence. Email mark.leake@york.ac.uk

**Supplementary Information includes:**

**9 Supplementary Figures**

**6 Supplementary Tables**

**7 Supplementary Movie Legends**

# Supplementary Figures

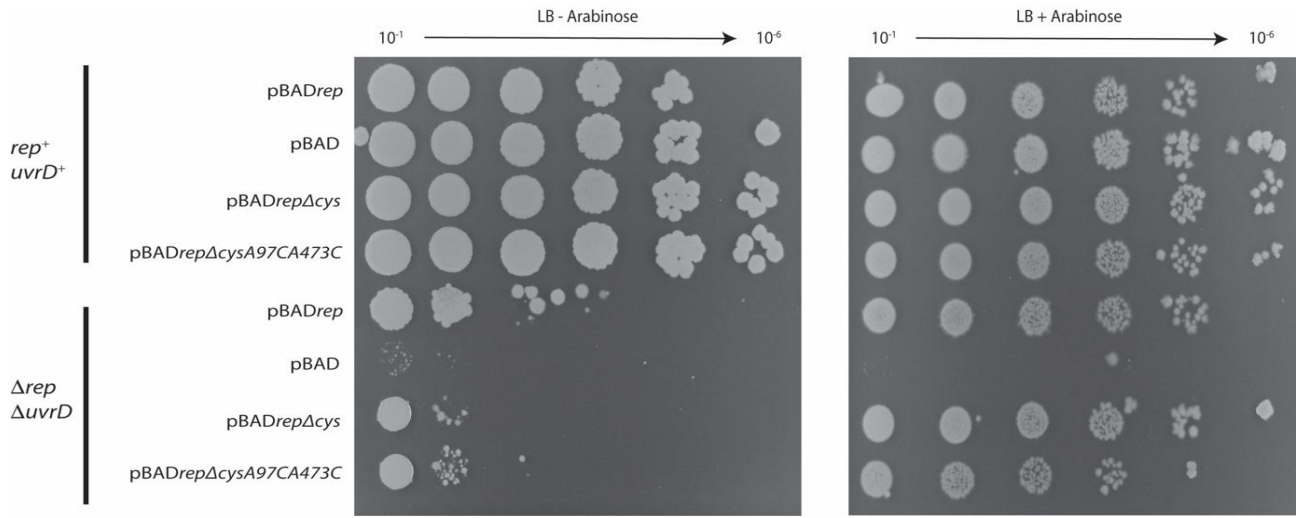

**Supplementary Figure 1. Complementation assays reveal that *repΔcysA97CA473C* is functional.** Two complementation assays are shown in the absence (left) and presence (right) of arabinose which the synthetic lethality of a *Δrep ΔuvrD* can be rescued by a covering plasmid containing *repΔcysA97CA473C*. When grown on rich media the *Δrep ΔuvrD* is lethal due to conflicts in DNA replication (see lack of growth in *Δrep ΔuvrD* pBAD spots). This phenotype can be rescued by a covering plasmid overexpressing Rep, growth is restored to normal level in *Δrep ΔuvrD* pBAD*rep* in the presence of arabinose. This synthetic lethality is also rescued by covering plasmids overexpressing RepΔcys or RepΔcysA97CA473C, growth is restored to normal levels in in *Δrep ΔuvrD* pBAD*repΔcys* and in *Δrep ΔuvrD* pBAD*repΔcysA97CA473C*.

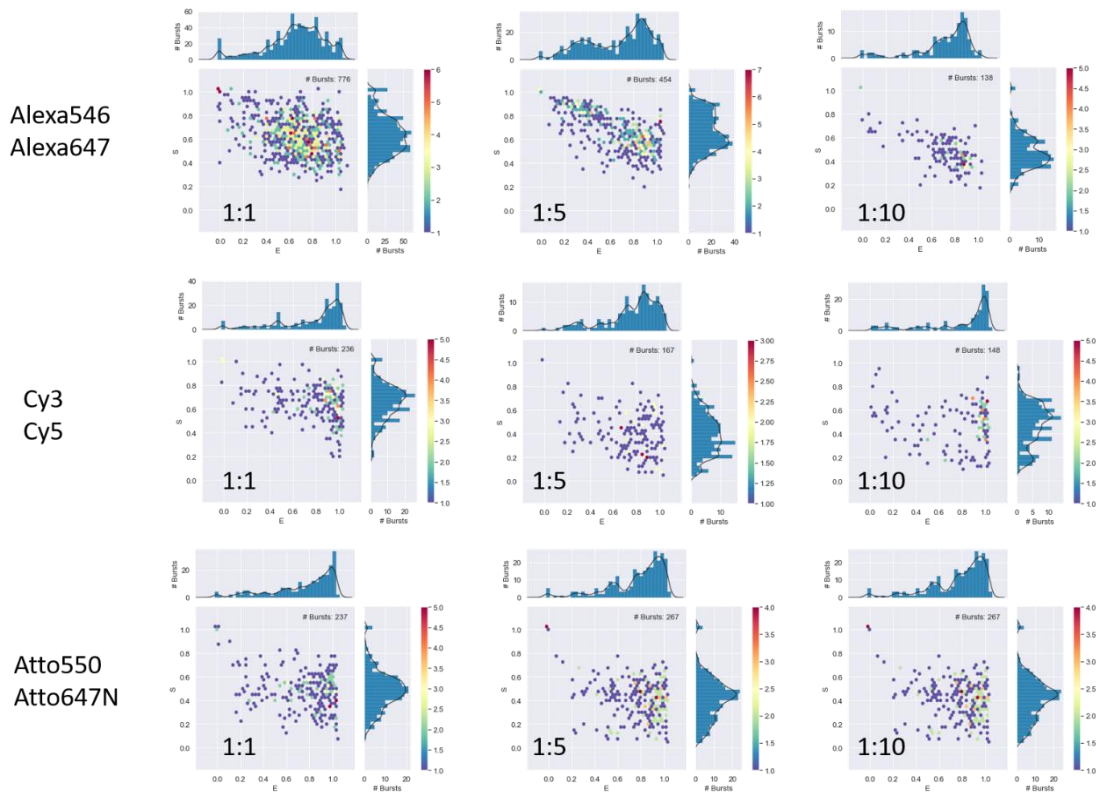

**Supplementary Figure 2. Different FRET dye pairs were triaged using confocal smFRET**. Using alternating laser excitation, we characterised the E-S relationships of several candidate FRET dye pairs over a range of relative concentrations, indicating that using Alexa Fluor 546 and Alexa Fluor 647 as a dye pair in a 1:1 labelling ration yielded the largest number of bursts using confocal smFRET as well as showing predominantly species containing both a donor and acceptor dye (stoichiometry parameter S= ~0.5, top left plot). Other dyes yielded fewer bursts for a given protein concentration, suggesting either poorer labelling efficiency or the presence of larger aggregates that are not seen in confocal smFRET. It is also clear that other dye pairs and dye ratios led to more donor or acceptor only bursts (S is approximately 1 or 0 respectively), which is sub optimal for smFRET measurements.

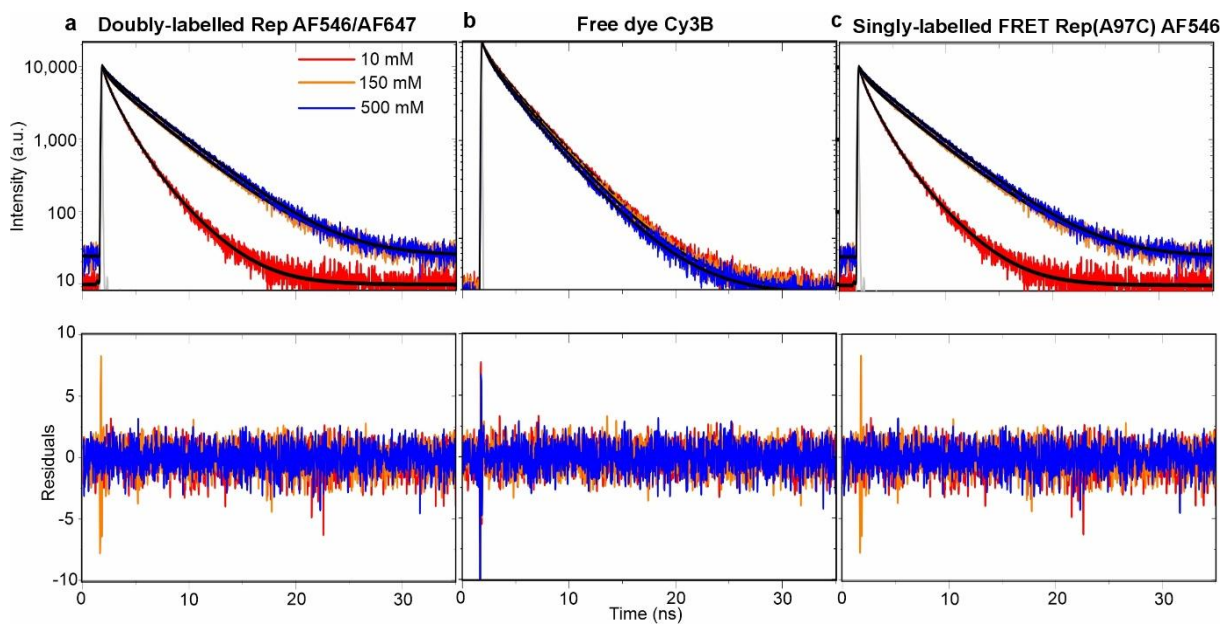

**Supplementary Figure 3. Ensemble FRET spectroscopy demonstrates dependence of FRET lifetime on NaCl concentration**. Fluorescence intensity decay profiles of **a.** doubly-labelled rep, **b.** Cy3B free dye, **c.** Rep labelled at position 97C with Alexa Fluor 546 under 10 mM (red), 150 mM (orange) and 500 mM NaCl (blue) conditions. Solid black lines represent tri-exponential fits to the data and numbers (insets) are the amplitude weighted average lifetimes obtained under each condition. Intensity decays were obtained with excitation laser wavelength = 532 nm and emission detection peak wavelength = 572 nm under magic angle conditions. Number of technical replicates n = 3.

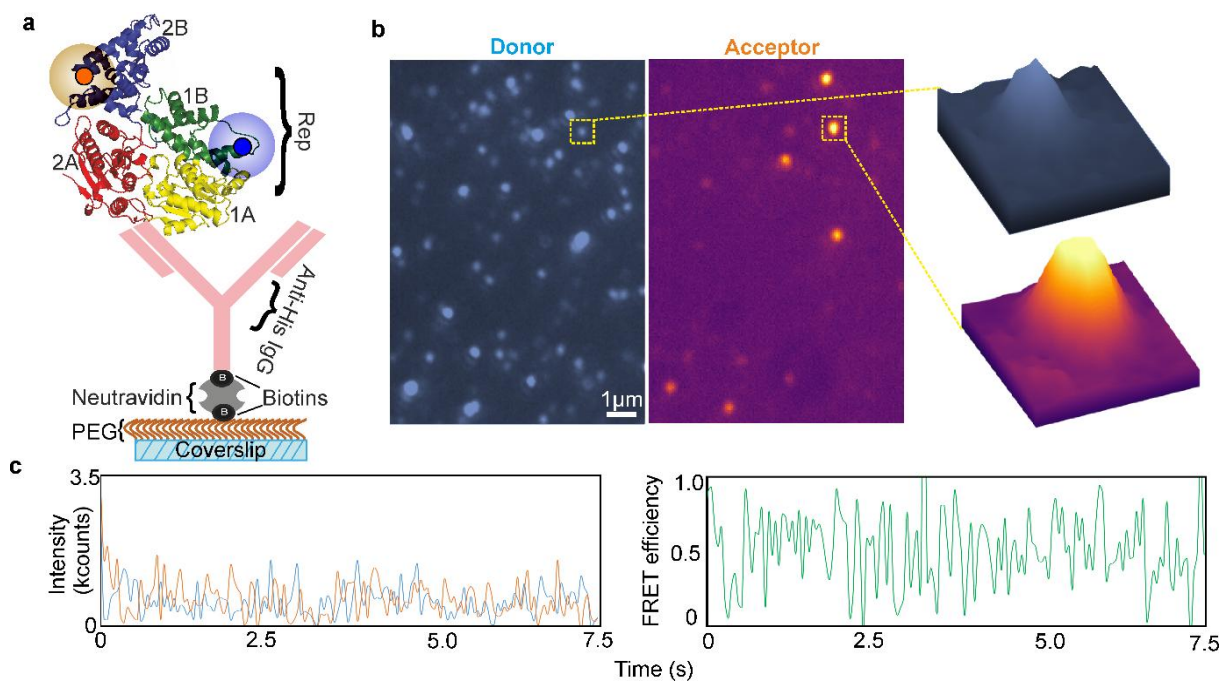

**Supplementary Figure 4. Single-molecule fluorescence microscopy of surface-immobilized Rep shows impairment of Rep activity due to the closeness of the surface**. **a.** Schematic of the immobilization of a single doubly labelled Rep conjugated via PEG-biotin/Neutravidin and Anti-His to the 6x His tag on Rep's N-terminus. It should be noted that an IgG is quite large with a typical Stokes radius of ~5 nm, so since Rep would then be typically at least ~10 nm from the surface there could still be some level of protein mobility within the TIRF evanescent field which could contribute to potential apparent noise. **b.** Representative widefield TIRF image of surface-immobilized Rep emitting in both donor (left) and acceptor (right) detector channels during donor-only excitation. Insets: zoom-ins of single immobilized foci. **c.** Representative donor (blur) and acceptor (orange) emission (left) with associated FRET efficiency (green, right) of a single Rep molecule displaying dynamic anti-correlated FRET behaviour.

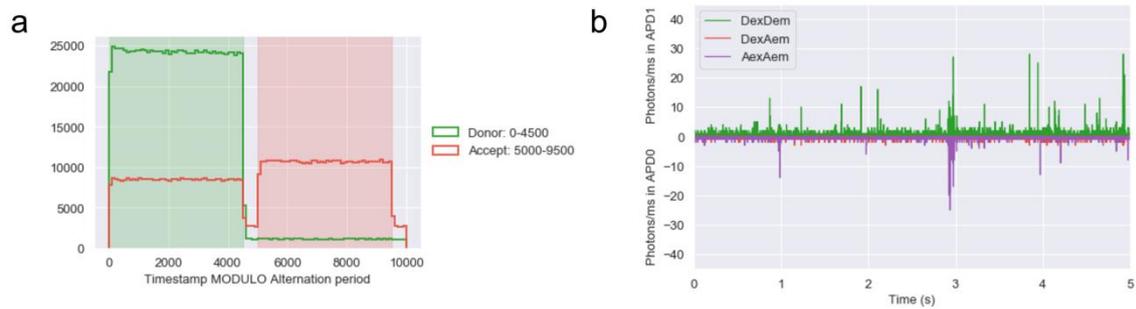

**Supplementary Figure 5. Rapid µs alternating laser excitation (ALEX) can be used for excitation of the donor and acceptor separately. a.** Histogram of photons by their arrival time within the ALEX cycle. The green laser is on during the green shaded area, and the red laser is on during the red shaded area. Green and red lines represent photons detected by the donor and acceptor channels. **b.** Representative example of raw µs data from confocal smFRET showing the donor and detector channels reconciled into coincident ms-timescale bursts.

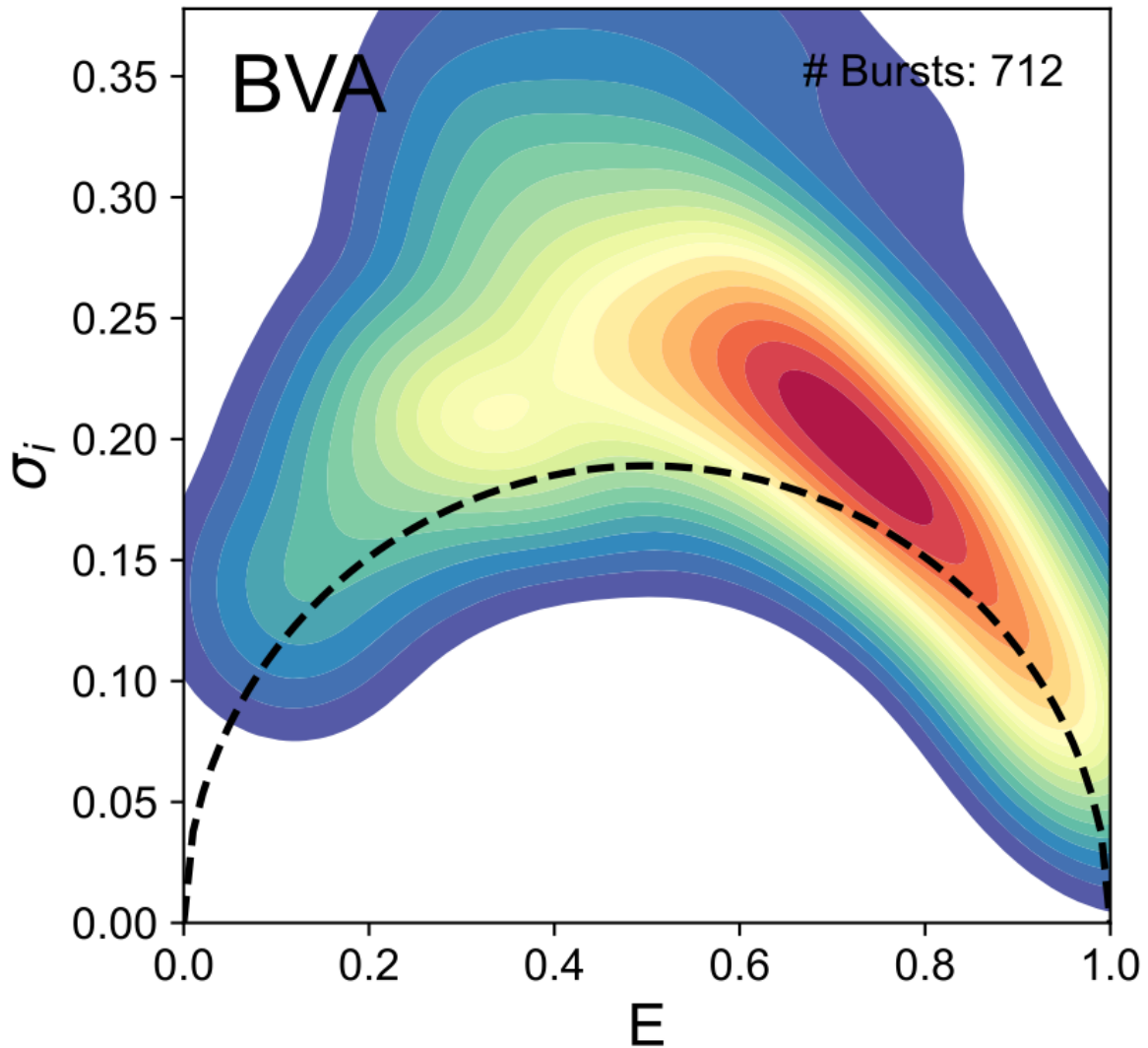

**Supplementary Figure 6. Burst variance analysis.** Heatmap showing burst variance analysis (BVA) on the low 10 mM NaCl data in the absence of DNA indicates that the bulk of the heatmap distribution of the SD of the inter-burst photon detection time ($\sigma_i$) was significantly high than theoretical expectations (black dashed line) for a system which does not undergo interconversion transitional kinetics[69], motivating subsequent quantitative H2MM analysis. Number of bursts = 712 obtained from low salt data.

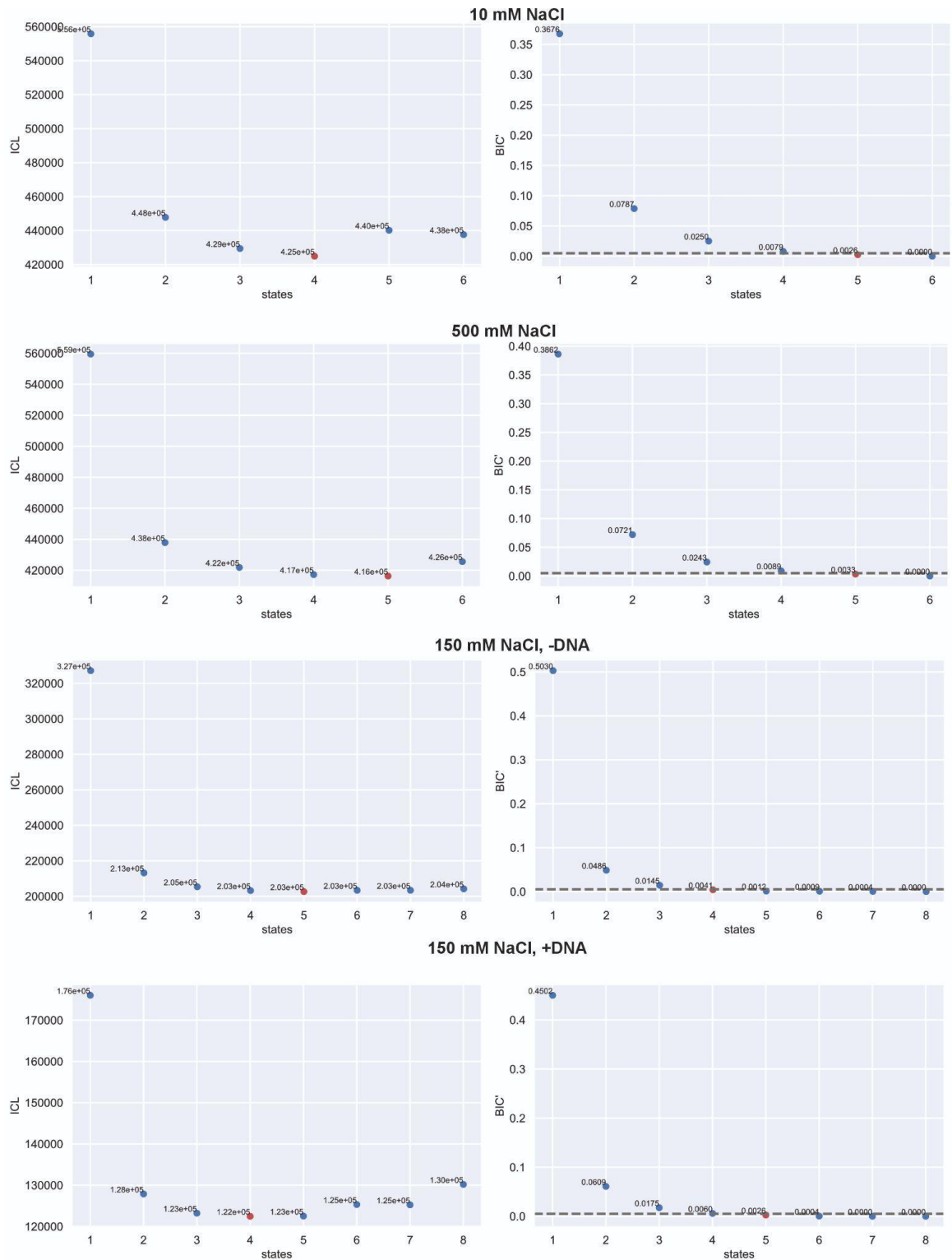

**Supplementary Figure 7. Comparison of ICL and BIC' for H2MM.** Plots indicating two different independent information criteria: ICL and modified BIC (BIC') as a function of the number of states in the H2MM models calculated (blue points). ICL results in a local minimum at the optimum model (red) for four or five states; further complexity from additional states is penalised and results in an increase. BIC' monotonically decreases and the optimal model is that which reaches a probability threshold of 0.005 (dashed line).

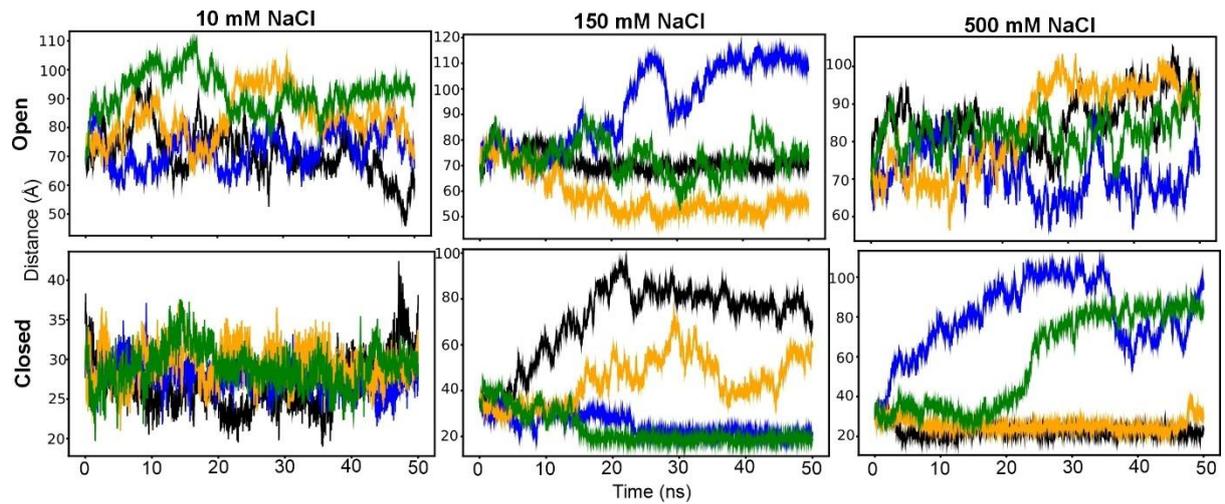

**Supplementary Figure 8. MD simulation predictions of distance between FRET-labelled protein residues indicate open-closed state transitions in the absence of DNA.** Time evolution of distance between residues 97 and 473 for each of the four replicates (different coloured lines) under each of the six conditions (low/intermediate/high salt and starting from open/closed conformations). Simulations initiated at the closed state and conducted at high salt were the only ones that do not show structural heterogeneity.

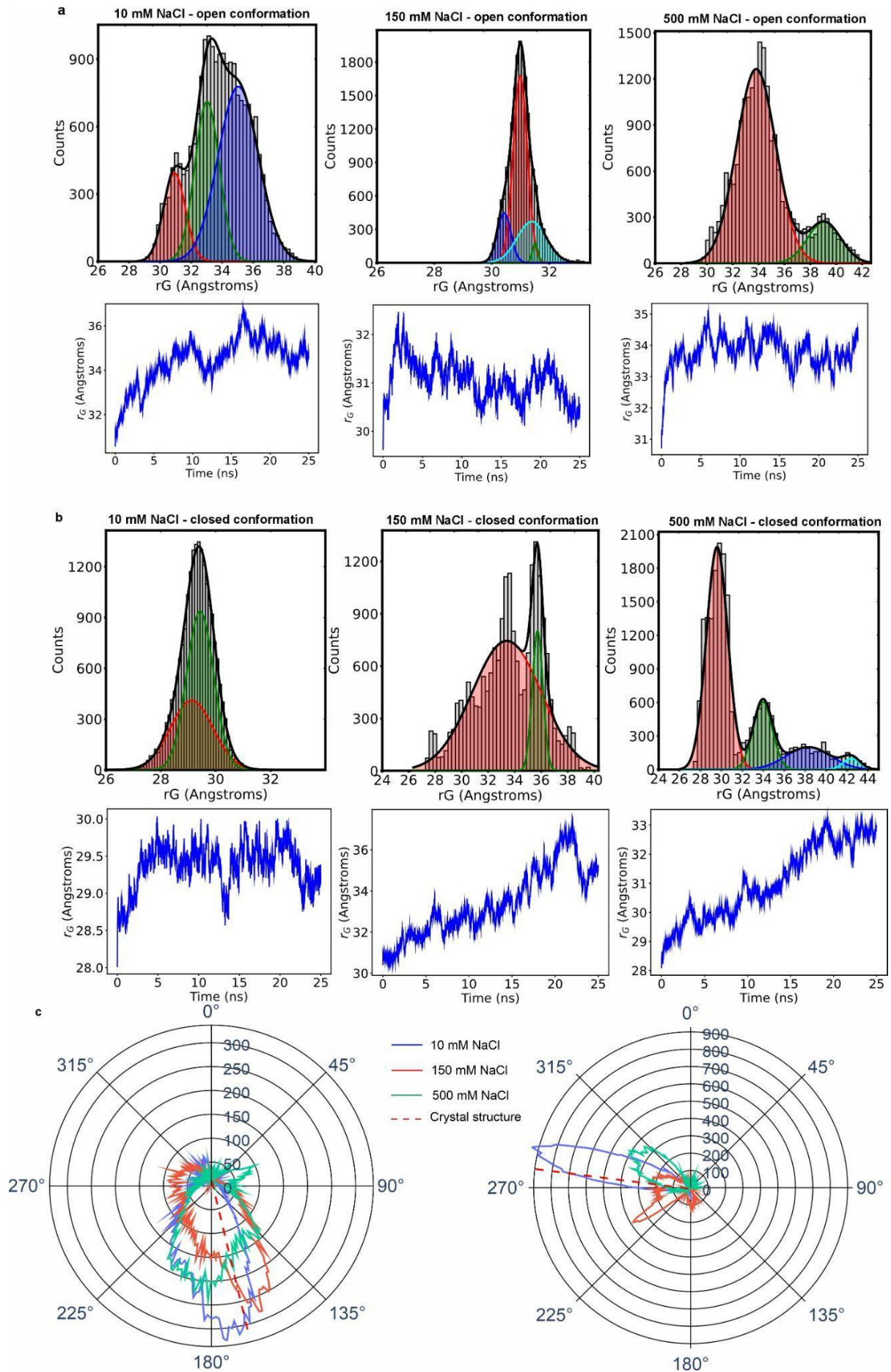

**Supplementary Figure 9. MDS predictions of free Rep radius of gyration and hinge rotation indicate structural heterogeneity.** Molecular dynamics simulations starting either from **a.** open or **b.** closed Rep structures indicate a distribution of time-dependent radius gyration (rG) states whose distributions can be fitted with between 2-4 Gaussian functions (goodness of fit $R^2$ ≥ 0.93 across all fits) depending on the level of NaCl concentration. Note, these Gaussian fits are performed free from any details of FRET levels. **c.** Population radar plots of the hinge rotation angle associated with WT-Rep initially in the open and closed conformations under 10 mM, 150 mM and 500 mM NaCl condition.

**Supplementary Tables**

| Strain name | Gene and plasmid details | Reference |
|---|---|---|
| N6524 | pAM403 (lac$^+$ rep$^+$) / ΔlacIZYA::<> | [11] |
| N6556 | pAM403 (lac$^+$ rep$^+$) / ΔlacIZYA::<> ΔuvrD::dhfr Δrep::cat | [11] |
| HB222 | E. coli B F$^-$ ompT hsdS$_B$(r$_B^-$ m$_B^-$) gal dcm lon [malB$^+$]$_{K12}$(λ$^S$) araB::T7RNAP-tetA Δrep::cat | [33] |

**Supplementary Table 1**. *Escherichia coli* K12 strains and plasmids used in this study.

| Optimised fitting parameters | NaCl 10 mM | NaCl 500 mM | NaCl 10 mM | NaCl 500 mM | NaCl 10 mM | NaCl 500 mM |
|---|---|---|---|---|---|---|
| | 1-component model: | 1-component model: | 2-component model: | 2-component model: | 3-component model: | 3-component model: |
| $a_1$ | 5.81 ± 0.80 | 8.76 ± 0.14 | 1.17 ± 0.02 | 3.18 ± 0.09 | 0.79 ± 0.02 | 2.08 ± 0.14 |
| $\tau_1$ (ns) | 1.34 ± 0.15 | 3.48 ± 0.03 | 2.99 ± 0.03 | 0.78 ± 0.03 | 3.46 ± 0.02 | 1.58 ± 0.33 |
| $a_2$ | -- | -- | 9.15 ± 0.26 | 7.34 ± 0.04 | 5.15 ± 0.16 | 6.78 ± 0.28 |
| $\tau_2$ (ns) | -- | -- | 0.56 ± 0.01 | 3.71 ± 0.01 | 0.87 ± 0.02 | 3.81 ± 0.04 |
| $a_3$ | -- | -- | -- | -- | 8.26 ± 0.49 | 2.52 ± 0.12 |
| $\tau_2$ (ns) | -- | -- | -- | -- | 0.26 ± 0.02 | 0.35 ± 0.08 |
| $\tau_{av}$ (ns) | 1.34 ± 0.15 | 3.48 ± 0.03 | 1.55 ± 0.01 | 3.47 ± 0.01 | 1.49 ± 0.01 | 3.47 ± 0.01 |
| Reduced $\chi^2$ | 36.4 | 4.88 | 2.29 | 1.27 | 1.42 | 1.17 |

**Supplementary Table 2** Fitting parameters corresponding to mono, bi- and tri-exponential models used to evaluate the fluorescence lifetime of doubly labelled Rep in low and high salt buffers. Errors are reported as errors of the fit. Convergence of the reduced $\chi^2$ parameter was used to evaluate that the fluorescence decays exhibited tri-, rather than mono- or bi-exponential behaviour.

| Optimised fitting parameters | 10 mM NaCl | 500 mM NaCl | 150 mM NaCl | 150 mM NaCl + DNA |
|---|---|---|---|---|
| 4-component model: | | | | |
| **Weight 1** | 11.0 | 24.5 | 18.0 | 13.8 |
| **Weight 2** | 23.3 | 24.9 | 15.5 | 22.2 |
| **Weight 3** | 47.4 | 33.2 | 30.3 | 31.4 |
| **Weight 4** | 18.3 | 17.4 | 36.1 | 32.6 |
| $<E_1> \pm \sigma$ | 0.21 ± 0.18 | 0.21 ± 0.13 | 0.11 ± 0.12 | 0.11 ± 0.11 |
| $<E_2> \pm \sigma$ | 0.54 ± 0.12 | 0.54 ± 0.12 | 0.44 ± 0.14 | 0.44 ± 0.14 |
| $<E_3> \pm \sigma$ | 0.82 ± 0.09 | 0.82 ± 0.09 | 0.8 ± 0.11 | 0.8 ± 0.10 |
| $<E_4> \pm \sigma$ | 0.98 ± 0.02 | 0.98 ± 0.02 | 0.97 ± 0.03 | 0.97 ± 0.03 |
| Goodness of fit $R^2$ | 0.91 | 0.88 | 0.77 | 0.80 |
| Reduced $\chi 2$ | 1.89 | | 3.21 | |
| BIC | -234.59 | | -354.34 | |
| 3-component model: | | | | |
| $<E_1> \pm \sigma$ | 0.23 ± 0.19 | 0.23 ± 0.14 | 0.23 ± 0.19 | 0.23 ± 0.18 |
| $<E_2> \pm \sigma$ | 0.60 ± 0.15 | 0.60 ± 0.13 | 0.76 ± 0.12 | 0.76 ± 0.14 |
| $<E_3> \pm \sigma$ | 0.89 ± 0.09 | 0.89 ± 0.08 | 0.97 ± 0.03 | 0.97 ± 0.03 |
| Goodness of fit $R^2$ | 0.66 | 0.52 | 0.76 | 0.78 |
| Reduced $\chi 2$ | 6.51 | | 4.46 | |
| BIC | -82.98 | | -292.96 | |
| 2-component model: | | | | |
| $<E_1> \pm \sigma$ | 0.42 ± 0.23 | 0.42 ± 0.23 | 0.47 ± 0.30 | 0.47 ± 0.28 |
| $<E_2> \pm \sigma$ | 0.87 ± 0.11 | 0.87 ± 0.10 | 0.95 ± 0.05 | 0.95 ± 0.05 |
| Goodness of fit $R^2$ | 0.73 | 0.59 | 0.63 | 0.67 |
| Reduced $\chi 2$ | 7.40 | | 7.75 | |
| BIC | -51.77 | | -157.61 | |

**Supplementary Table 3** Optimized parameters for Gaussian fits for smFRET using confocal microscopy on labelled Rep in low (10 mM) and high (500 mM) NaCl as well as an intermediate NaCl concentration (150 mM) ± DNA. Taken using number of bursts n = 1,021 and 1,013 from high and low salt respectively, from 7 different samples, for n = 1,141 from 9 different samples for intermediate salt, and for n = 1250 from 10 different samples for intermediate salt + DNA. $<E_i>$ value is the mean FRET efficiency values for the i[th] Gaussian curve in the fit. Corresponding mean FRET efficiencies with $R^2$, reduced Chi squared and BIC metrics shown for 4-, 3- and 2-component models, all of which show that the 2- and 3-component models clearly exhibit worse fits than the 4-component model. The addition of a 5[th] component resulted in convergence to one of the existing 4 components and demonstrated a less negative BIC metric than the 4-componet model, thus supporting a 4-component best fit model.

|  | 10 mM NaCl | | 500 mM NaCl | | 150 mM NaCl | | 150 mM NaCl + DNA | |
|---|---|---|---|---|---|---|---|---|
|  | FRET Efficiency | Relative Occupancy | FRET Efficiency | Relative Occupancy | FRET Efficiency | Relative Occupancy | FRET Efficiency | Relative Occupancy |
| S1 | 0.2050 ± 0.0031 | 35.72% | 0.2067 ± 0.0021 | 32.28% | 0.2388 ± 0.0038 | 17.41% | 0.2696 ± 0.0046 | 25.84% |
| S2 | 0.5705 ± 0.0036 | 31.95% | 0.4911 ± 0.0034 | 23.28% | 0.5114 ± 0.0059 | 26.29% | 0.5648 ± 0.0065 | 24.27% |
| S3 | 0.7292 ± 0.0021 | 22.72% | 0.7212 ± 0.0025 | 27.23% | 0.8093 ± 0.0038 | 29.66% | 0.8108 ± 0.0048 | 25.69% |
| S4 | 0.9454 ± 0.0013 | 9.61% | 0.9441 ± 0.0013 | 17.21% | 0.9599 ± 0.0015 | 26.65% | 0.9585 ± 0.0017 | 24.20% |

| | | | Transition Rate ($s^{-1}$) [lower bound, upper bound] | | | |
|---|---|---|---|---|---|---|
| | | | End State | | | |
| | | | S1 | S2 | S3 | S4 |
| 10 mM NaCl | Start State | S1 |  | 29005.55 [28355.78, 29976.38] | 958.31 [918.36, 999.33] | 28.00 [13.39, 44.14] |
| | | S2 | 25069.40 [24479.10, 26015.52] |  | 5.19 [0.00, 35.25] | 0.00 [0.00, 8.02] |
| | | S3 | 390.08 [373.38, 407.38] | 28.71 [11.38, 44.77] |  | 306.27 [292.01, 320.98] |
| | | S4 | 0.00 [0.00, 4.54] | 0.00 [0.00, 2.46] | 463.88 [441.08, 487.27] |  |
| 500 mM NaCl | Start State | S1 |  | 951.63 [912.67, 991.60] | 0.00 [0.00, 6.67] | 74.28 [64.23, 84.93] |
| | | S2 | 1003.92 [959.82, 1049.31] |  | 367.50 [344.56, 391.18] | 0.00 [0.00, 4.83] |
| | | S3 | 14.70 [6.48, 23.98] | 344.98 [327.13, 363.50] |  | 312.05 [294.52, 330.17] |
| | | S4 | 33.95 [24.30, 44.54] | 0.00 [0.00, 9.07] | 437.46 [413.86, 461.81] |  |
| 150 mM NaCl | Start State | S1 |  | 778.04 [690.92, 871.28] | 17.03 [0.00, 63.33] | 0.00 [0.00, 16.42] |
| | | S2 | 943.71 [853.24, 1039.67] |  | 1296.72 [1186.22, 1412.58] | 0.00 [0.00, 31.13] |
| | | S3 | 0.00 [0.00, 21.69] | 1011.81 [929.22, 1098.82] |  | 611.48 [545.13, 681.60] |
| | | S4 | 0.00 [0.00, 5.62] | 0.00 [0.00, 9.79] | 544.14 [485.49, 606.10] |  |
| 150 mM NaCl + DNA | Start State | S1 |  | 1262.28 [1154.72, 1375.06] | 0.00 [0.00, 30.25] | 0.00 [0.00, 8.31] |
| | | S2 | 1572.31 [1440.58, 1711.40] |  | 945.39 [843.48, 1053.97] | 11.43 [0.00, 55.65] |
| | | S3 | 0.00 [0.00, 20.38] | 883.52 [793.22, 979.14] |  | 1096.49 [989.05, 1209.74] |
| | | S4 | 13.12 [0.00, 35.50] | 0.00 [0.00, 30.24] | 948.53 [861.62, 1039.51] |  |

**Supplementary Table 4** Optimised parameters from H2MM analysis. Upper panel shows the optimised outputs for the relative occupancy of the four states. Lower panel indicates the optimised interconversion rate constants between all four states. Errors and low/upper limits quoted are based on log-likelihood uncertainty estimation (see Methods).

| State interconversion transitions | Rep | | | | UvrD |
|---|---|---|---|---|---|
| | $k$ ($s^{-1}$) in 10mM NaCl | $k$ ($s^{-1}$) in 150mM NaCl, no DNA | $k$ ($s^{-1}$) in 150mM NaCl, +DNA | $k$ ($s^{-1}$) in 500 mM NaCl | $k$ ($s^{-1}$) in 60 mM NaCl |
| S1 to S2 | >>1,000 | 1,262 | 1,572 | 952 | 0.85 |
| S2 to S1 | >>1,000↔ | 945↔ | 1,262↔ | 1,004↔ | 0.26↓ |
| S2 to S3 | 5 | 1,297 | 945 | 368 | 0.30 |
| S3 to S2 | 29↑ | 1,021↔ | 884↔ | 345↔ | 0.25↓ |
| S3 to S4 | 306 | 611 | 1096 | 312 | |
| S4 to S3 | 464↔ | 544↔ | 948↔ | 437↔ | 0.20↓ |
| S1 to S3 | 958 | <1 | <1 | < 1 | 0.16 |
| S3 to S1 | 390↓ | <1 | <1 | 15↑ | 0.18↔ |
| S2 to S4 | <1 | <1 | 11↑ | <1 | 0.54 |
| S4 to S2 | <1↔ | <1 | <1 | <1↔ | 0.33↓ |
| S4 to S1 | <1 | <1 | 13↑ | 34 | -- |
| S1 to S4 | 28↑ | <1 | <1 | 74↑ | -- |

**Supplementary Table 5** Comparison of transitional kinetics rate constants for states previously identified for UvrD[23] and those measured in our study for Rep at the high and low levels. Yellow highlight indicates > 2x difference between rates at low (10 mM) and intermediate (150 mM) and/or high (500 mM) NaCl concentration for Rep in absence of DNA; ↓ indicates rate between states SY to SX relative to SX to SY decreases; ↔ indicates rates between SY to SX rate remains unchanged within a factor of 2 relative to rates between states SX to SY; ↑ indicates rate between states SY to SX increases relative to rate between states SX to SY. Rate constants at 150 mM NaCl in presence of DNA are also included for reference.

| Parameter | Alexa Fluor 546 (donor) | Alexa Fluor 647 (acceptor) |
|---|---|---|
| Linker length (Å) | 14 | 14 |
| Linker width (Å) | 4.5 | 4.5 |
| Radius 1 (Å) | 6.8 | 11 |
| Radius 2 (Å) | 3.9 | 3 |
| Radius 3 (Å) | 1.8 | 1.5 |
| $R_0$ (Å) | 68.2 | |

**Supplementary Table 6. Fluorophores with their respective linkers that were implemented onto modelling structures of Rep.**

**Supplementary movie screenshots and legends**

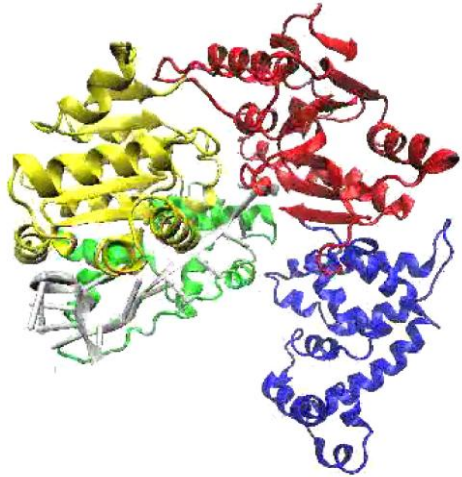

**Supplementary Movie 1** MD simulation (starting from open conformation), Rep with DNA bound, low salt (10 mM NaCl), implicit solvent.

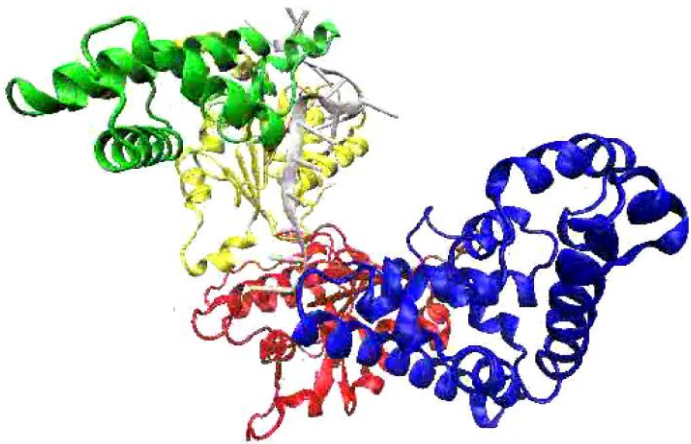

**Supplementary Movie 2** MD simulation (starting from open conformation), Rep with DNA bound, high salt (500 mM NaCl), implicit solvent.

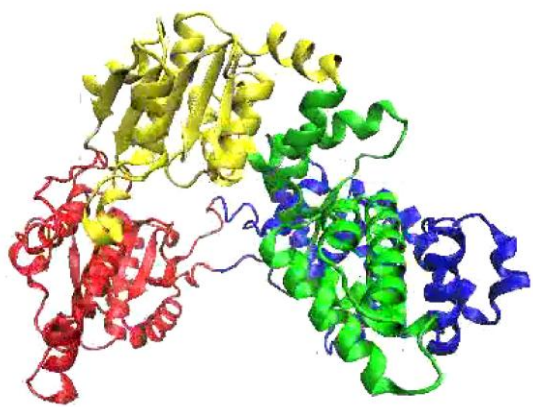

**Supplementary Movie 3** MD simulation (starting from open conformation), Rep with no DNA, low salt (10 mM NaCl), implicit solvent.

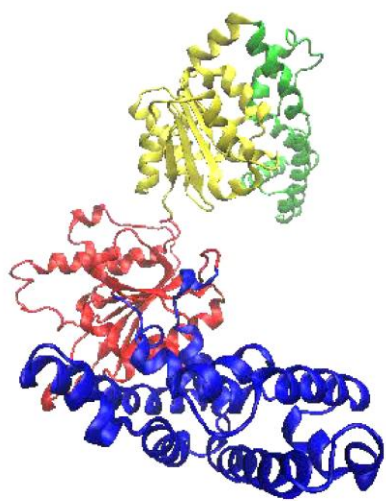

**Supplementary Movie 4** MD simulation (starting from open conformation), Rep with no DNA, high salt (500 mM NaCl), implicit solvent.

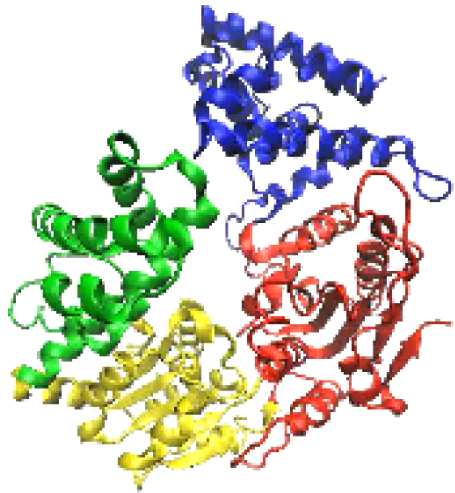

**Supplementary Movie 5** MD simulation (starting from open conformation), Rep with no DNA, intermediate salt (150 mM NaCl), implicit solvent.

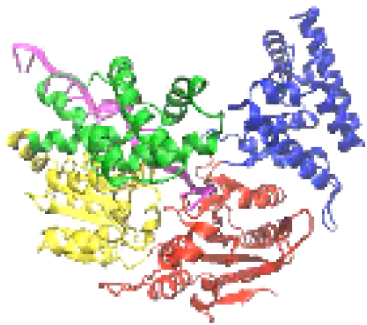

**Supplementary Movie 6** MD simulation (starting from open conformation), Rep with DNA (shown in magenta), intermediate salt (150 mM NaCl), implicit solvent.

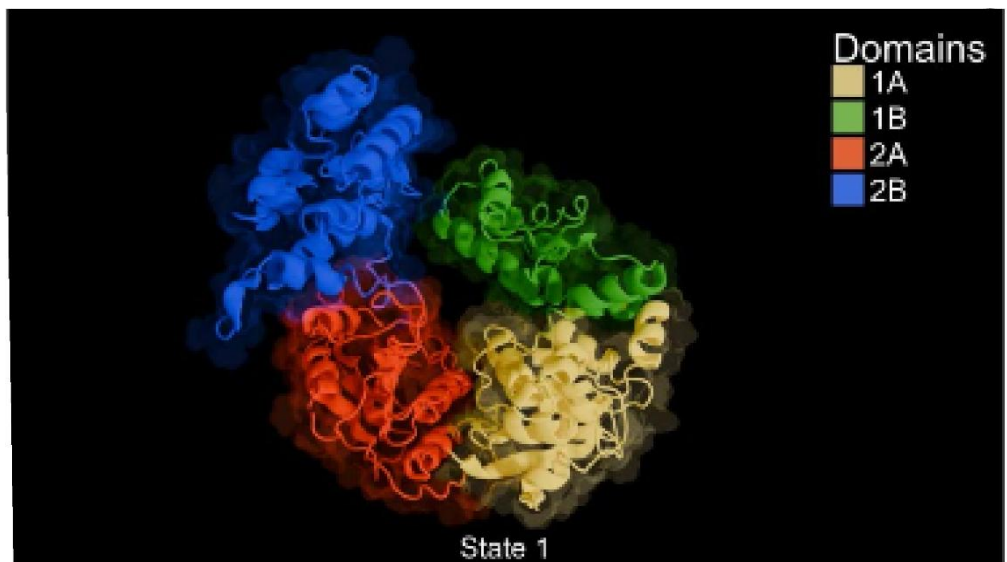
**Supplementary Movie 7** Animation depicting structural transitions between states S1, S2, S3 and S4.